\RequirePackage{rotating}
\documentclass[twocolumn, tighten]{aastex62}
\usepackage{amsmath}
\usepackage[caption=false]{subfig}
\usepackage{xfrac}

\shorttitle{DSHARP II: Annular substructures}
\shortauthors{Huang et al.}

\begin{document}

\title{The Disk Substructures at High Angular Resolution Project (DSHARP):\\
II. Characteristics of Annular Substructures}

\correspondingauthor{Jane Huang}
\email{jane.huang@cfa.harvard.edu}

\author{Jane Huang}
\affiliation{Harvard-Smithsonian Center for Astrophysics, 60 Garden Street, Cambridge, MA 02138, United States of America}

\author{Sean M. Andrews}
\affiliation{Harvard-Smithsonian Center for Astrophysics, 60 Garden Street, Cambridge, MA 02138, United States of America}

\author{Cornelis P. Dullemond}
\affiliation{Zentrum f\"ur Astronomie, Heidelberg University, Albert Ueberle Str. 2, 69120 Heidelberg, Germany}

\author{Andrea Isella}
\affiliation{Department of Physics and Astronomy, Rice University, 6100 Main Street, Houston, TX 77005, United States of America}

\author{Laura M. P\'erez}
\affiliation{Departamento de Astronom\'ia, Universidad de Chile, Camino El Observatorio 1515, Las Condes, Santiago, Chile}

\author{Viviana V. Guzm\'an}
\affiliation{Joint ALMA Observatory, Avenida Alonso de C\'ordova 3107, Vitacura, Santiago, Chile}
\affiliation{Instituto de Astrof\'isica, Pontificia Universidad Cat\'olica de Chile, Av. Vicu\~na Mackenna 4860, 7820436 Macul, Santiago, Chile}

\author{Karin I. \"Oberg}
\affiliation{Harvard-Smithsonian Center for Astrophysics, 60 Garden Street, Cambridge, MA 02138, United States of America}

\author{Zhaohuan Zhu}
\affiliation{Department of Physics and Astronomy, University of Nevada, Las Vegas, 4505 S. Maryland Pkwy, Las Vegas, NV, 89154, USA}

\author{Shangjia Zhang}
\affiliation{Department of Physics and Astronomy, University of Nevada, Las Vegas, 4505 S. Maryland Pkwy, Las Vegas, NV, 89154, USA}

\author{Xue-Ning Bai}
\affiliation{Institute for Advanced Study and Tsinghua Center for Astrophysics, Tsinghua University, Beijing 100084, China}

\author{Myriam Benisty}
\affiliation{Unidad Mixta Internacional Franco-Chilena de Astronom\'{i}a, CNRS/INSU UMI 3386, Departamento de Astronom\'ia, Universidad de Chile, Camino El Observatorio 1515, Las Condes, Santiago, Chile}
\affiliation{Univ. Grenoble Alpes, CNRS, IPAG, 38000 Grenoble, France}

\author{Tilman Birnstiel}
\affiliation{University Observatory, Faculty of Physics, Ludwig-Maximilians-Universit\"at M\"unchen, Scheinerstr. 1, 81679 Munich, Germany}

\author{John M. Carpenter}
\affiliation{Joint ALMA Observatory, Avenida Alonso de C\'ordova 3107, Vitacura, Santiago, Chile}

\author{A. Meredith Hughes}
\affiliation{Department of Astronomy, Van Vleck Observatory, Wesleyan University, 96 Foss Hill Drive, Middletown, CT 06459, USA}

\author{Luca Ricci}
\affiliation{Department of Physics and Astronomy, California State University Northridge, 18111 Nordhoff Street, Northridge, CA 91130, USA}

\author{Erik Weaver}
\affiliation{Department of Physics and Astronomy, Rice University, 6100 Main Street, Houston, TX 77005, United States of America}

\author{David J. Wilner}
\affiliation{Harvard-Smithsonian Center for Astrophysics, 60 Garden Street, Cambridge, MA 02138, United States of America}

\begin{abstract}
The Disk Substructures at High Angular Resolution Project used ALMA to map the 1.25 millimeter continuum of protoplanetary disks at a spatial resolution of $\sim5$ au. We present a systematic analysis of annular substructures in the 18 single-disk systems targeted in this survey. No dominant architecture emerges from this sample; instead, remarkably diverse morphologies are observed. Annular substructures can occur at virtually any radius where millimeter continuum emission is detected and range in widths from a few au to tens of au. Intensity ratios between gaps and adjacent rings range from near-unity to just a few percent. In a minority of cases, annular substructures co-exist with other types of substructures, including spiral arms (3/18) and crescent-like azimuthal asymmetries (2/18). No clear trend is observed between the positions of the substructures and stellar host properties. In particular, the absence of an obvious association with stellar host luminosity (and hence the disk thermal structure) suggests that substructures do not occur preferentially near major molecular snowlines. Annular substructures like those observed in DSHARP have long been hypothesized to be due to planet-disk interactions. A few disks exhibit characteristics particularly suggestive of this scenario, including substructures in possible mean-motion resonance and ``double gap'' features reminiscent of hydrodynamical simulations of multiple gaps opened by a planet in a low-viscosity disk. 
\end{abstract}


\section{Introduction} \label{sec:intro}
Exoplanet surveys have uncovered remarkable diversity in the masses, radii, and composition of exoplanets, as well as in the overall architectures of exoplanetary systems \citep[e.g.,][]{2010Sci...330..653H, 2011ApJ...736...19B, 2015ARAA..53..409W, 2017AJ....154..109F}. These wide variations in properties are thought to be closely tied to differences in where and when planets formed in the protoplanetary disks orbiting young stars \citep[e.g.,][]{2011ApJ...743L..16O,2012AA...547A.111M, 2016ApJ...817...90L}. Recent discoveries of complex features in protoplanetary disks, such as annular substructures, spiral arms, and high-contrast azimuthal asymmetries, have fueled interest in using observed disk morphologies to infer the properties of an unseen population of newly-forming planets \citep[e.g.,][]{2012ApJ...748L..22M, 2013Sci...340.1199V,2015ApJ...808L...3A,2016PhRvL.117y1101I}. 

Of these various types of small-scale features, annular substructures have been reported most often. They have predominantly been detected in millimeter continuum emission tracing larger ($\sim$millimeter-sized) dust grains in the disk midplane \citep[e.g.,][]{2016ApJ...820L..40A,2018AA...610A..24F,2018ApJ...869...17L}, but have also been observed in scattered light tracing sub-micron-sized dust grains in the disk atmosphere \citep[e.g.,][]{2017ApJ...837..132V, 2018ApJ...863...44A,2018AA...614A..24M} and in molecular emission \citep[e.g.,][]{2016PhRvL.117y1101I, 2017AA...600A..72F, 2017ApJ...835..228T}. The detections of these annular substructures are particularly intriguing with respect to planet formation theory because numerical simulations have long predicted that a sufficiently massive protoplanet will trigger density waves that eventually shock and create disk gaps \citep[e.g.,][]{1984ApJ...285..818P,1999ApJ...514..344B,2000MNRAS.318...18N}. 

However, it has been unclear whether annular substructures (or any other kind of substructure, for that matter) necessarily arise from planet-disk interactions, and a number of other hypotheses have been proposed to explain their origins. One set of hypotheses suggests that condensation of certain molecular species may affect how dust grains fragment, grow, and drift, in turn causing dust to accumulate in proximity to various snowlines \citep[e.g.,][]{2015ApJ...815L..15B, 2015ApJ...806L...7Z, 2016ApJ...821...82O,2017ApJ...845...68P}. Another group of proposed mechanisms explores the effects of coupling between magnetic fields and disk material, such as zonal flows arising from magnetorotational instability (MRI) turbulence \citep[e.g.,][]{2009ApJ...697.1269J,2014ApJ...796...31B,2014ApJ...784...15S}, surface density enhancements at the edges of dead zones \citep[e.g.,][]{2015AA...574A..10L, 2015AA...574A..68F, 2016AA...590A..17R}, and the spontaneous concentration of magnetic flux \citep[e.g.,][]{2014ApJ...796...31B,2017AA...600A..75B,2018MNRAS.477.1239S}. Other proposed mechanisms have focused on instabilities arising from interactions between gas and dust, including secular gravitational instabilities resulting from friction between gas and dust \citep[e.g.,][]{2014ApJ...794...55T} or dust-driven viscous ring-instability resulting from dust drift altering the viscosity profile of the disk and amplifying density perturbations \citep{2018AA...609A..50D}. 

While much of our current knowledge of disk substructure comes from serendipitously discovered case studies, establishing the origins of annular substructures in disks can be facilitated by a systematic analysis of their properties. To that end, we undertook the Disk Substructures at High Angular Resolution Project (DSHARP), the first high angular resolution ALMA survey of disks \citep{2018ApJ...869L..41A}. Substructures are detected in the 1.25 millimeter continuum emission of all 20 sources, lending weight to the notion that millimeter continuum substructures are common and perhaps even ubiquitous. In this work, we present an analysis of the annular substructures observed in single-disk systems. Analysis of the two multiple-disk systems in DSHARP is presented in \citet{2018ApJ...869L..44K}.  Section \ref{sec:overview} provides a brief overview of the disk sample. In Section \ref{sec:features}, we measure the locations, widths, and contrasts of the substructures. In Section \ref{sec:trends}, we examine trends in their positions and sizes. Section \ref{sec:discussion} comments on possible origins for the substructures and discusses implications for their prevalence in the general disk population. Section \ref{sec:summary} summarizes the main findings. 
\section{Sample overview\label{sec:overview}}
DSHARP targeted 20 systems, 18 of which are single-disk systems (i.e., all sources except AS 205 and HT Lup). In the context of this paper, ``the DSHARP sample'' refers specifically to these 18 disks. Angular resolutions range from $\sim30$ to 60 mas, probing scales of $\sim5$ to 8 au. Stellar properties, including distances, masses, accretion rates, ages, and luminosities, are listed in Table 1 of \citet{2018ApJ...869L..41A}. The analysis in this work is based on the ``fiducial'' images presented in \citet{2018ApJ...869L..41A}, which also provides details about the observational setup and the calibration and imaging procedure. Table 4 of \citet{2018ApJ...869L..41A}  provides basic information about these images, including synthesized beam sizes, peak intensities, continuum flux densities, and image rms noise levels. 

Some portions of the analysis fold in publicly available continuum observations of HL Tau and TW Hya taken at comparable spatial resolution and sensitivity. The TW Hya analysis uses the 1.0 mm image presented in \citet{2018ApJ...852..122H} and the HL Tau analysis uses the 1.0 mm image presented in \citet{2015ApJ...808L...3A}. Although 1.3 mm continuum observations are available for both sources \citep{2015ApJ...808L...3A,2016ApJ...829L..35T}, the 1.0 mm images are used instead because of their better angular resolution and $uv$ coverage. 

\begin{figure*}
\begin{center}
\includegraphics{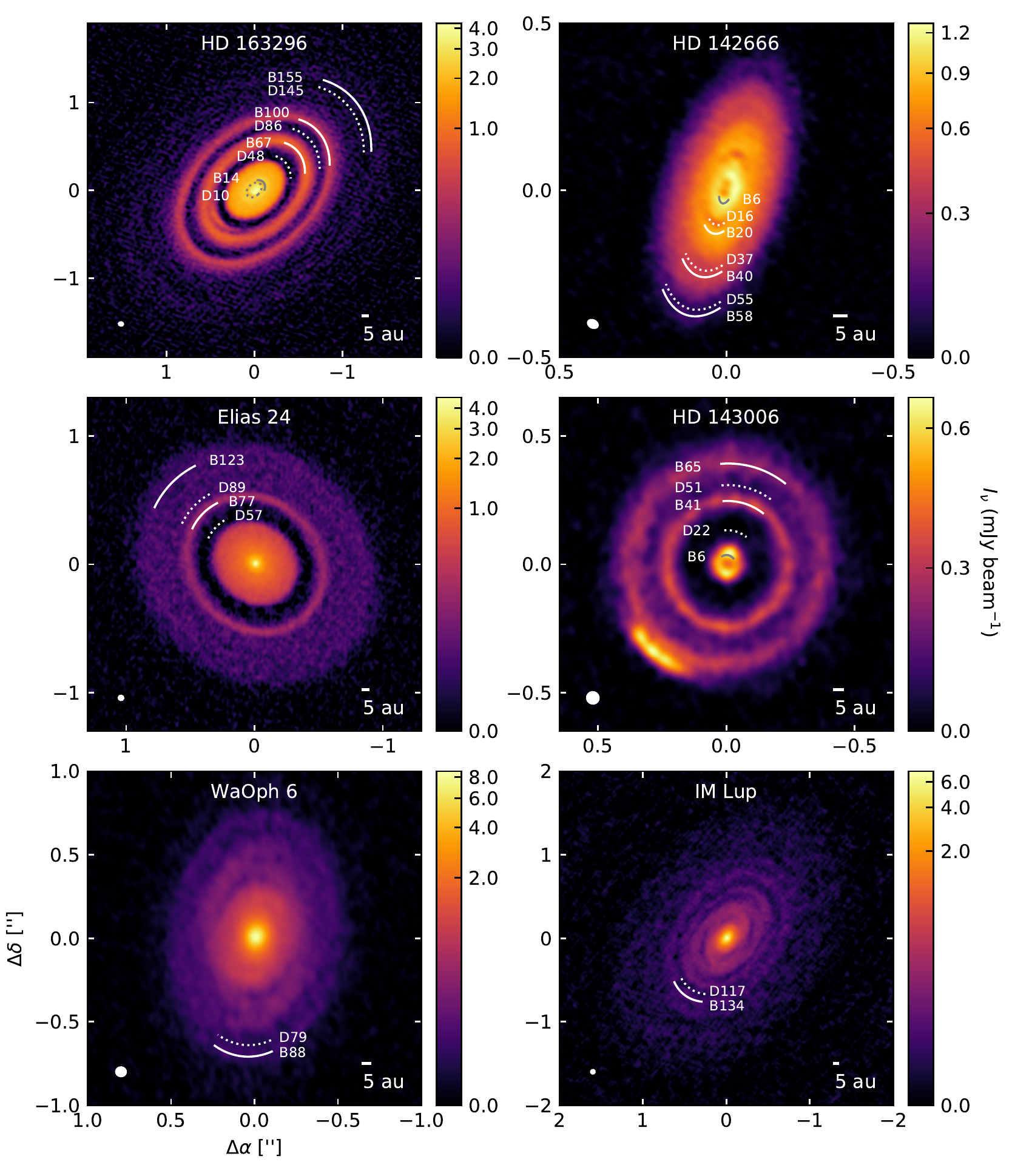}
\end{center}
\caption{ALMA 1.25 mm continuum images of the DSHARP sample, ordered by decreasing stellar luminosity (see Table 1 of \citealt{2018ApJ...869L..41A}) from left to right and top to bottom. An arcsinh stretch is applied to the color scale of each disk to increase the visibility of substructures in the outer regions. Axes are labeled with angular offsets from the disk center. Annular substructures are marked with dotted arcs (gaps) or solid arcs (bright emission rings). The synthesized beam is shown in the lower left corner of each panel. \label{fig:annotatedrings}}
\end{figure*}

\begin{figure*}
\begin{center}
\ContinuedFloat
\includegraphics{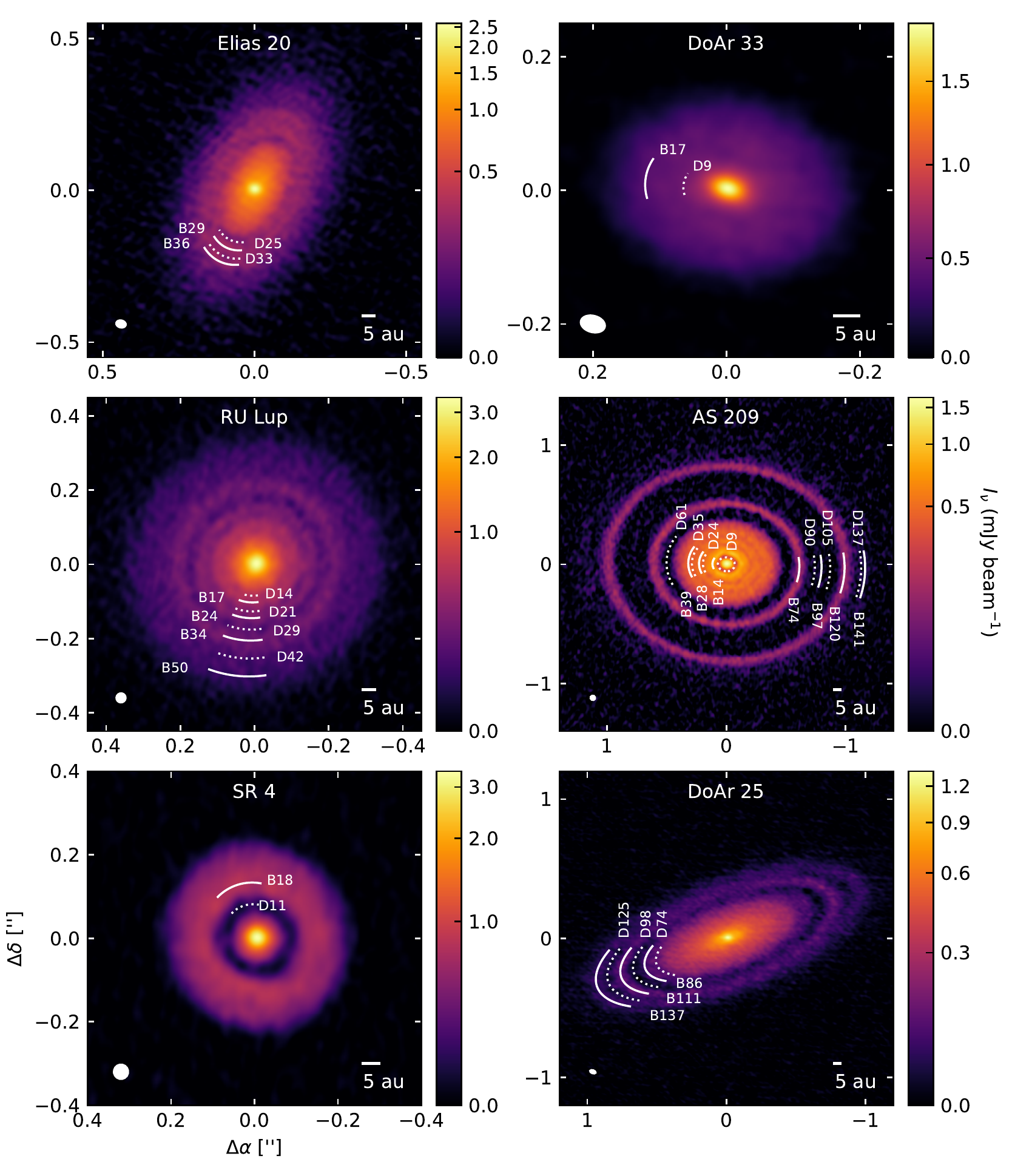}
\end{center}
\caption{Continued}
\end{figure*}

\begin{figure*}
\begin{center}
\ContinuedFloat
\includegraphics{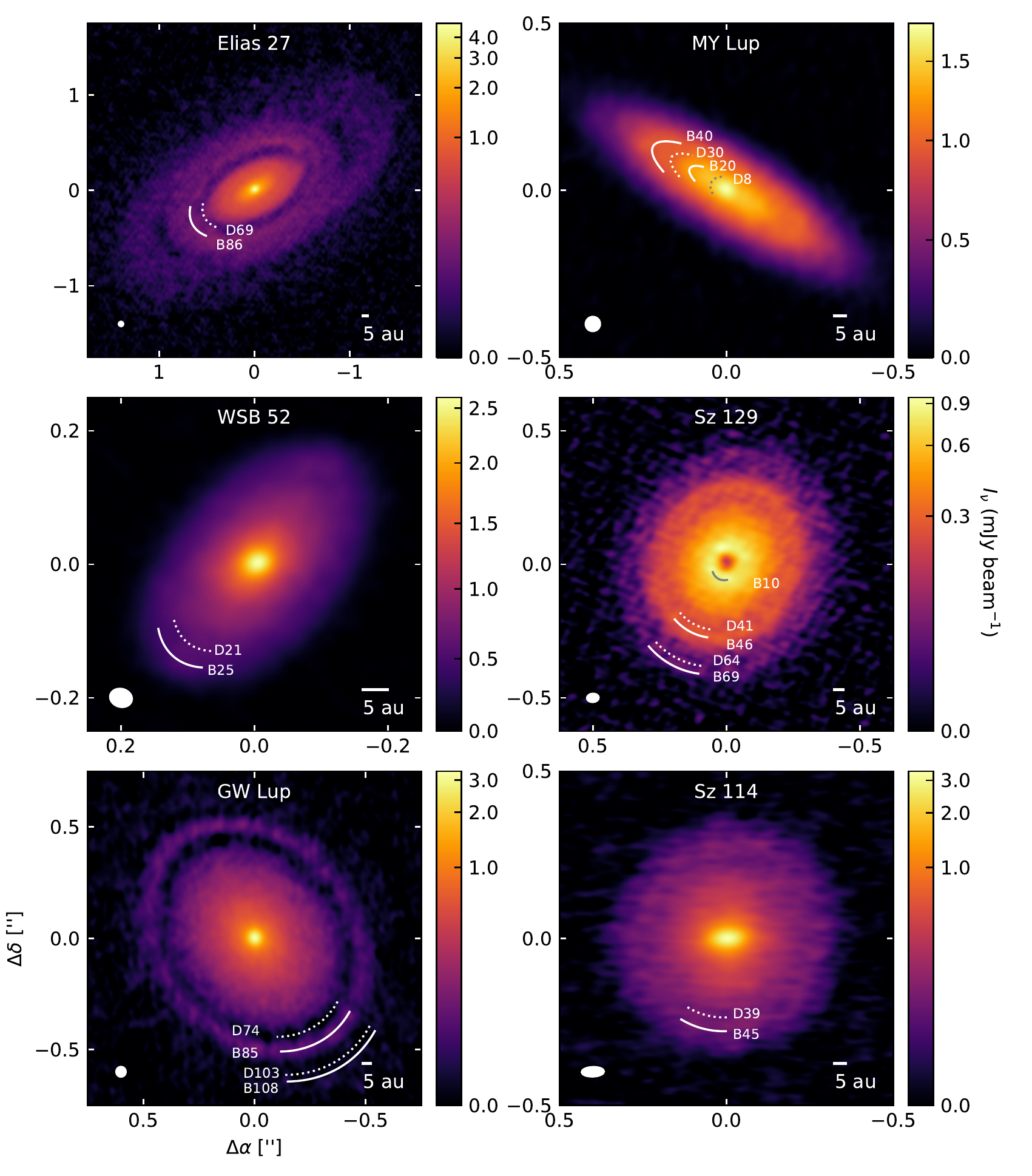}
\end{center}
\caption{Continued}
\end{figure*}

\section{Disk features}\label{sec:features}
\subsection{Radial locations of annular substructures}
We use the term ``annular substructures'' to refer collectively to the ``bright'' and ``dark'' annuli visible in the DSHARP 1.25 mm continuum images. Colloquially, the ``bright'' and ``dark'' features are often referred to as ``rings'' and ``gaps,'' respectively \citep[e.g.,][]{2015ApJ...808L...3A,2016PhRvL.117y1101I, 2017AA...600A..72F}.  Figure \ref{fig:annotatedrings} shows the 1.25 mm continuum images of the DSHARP sources with labeled annular substructures. The three-part procedure used to identify and label these substructures is explained below.

First, many of the emission features manifest as well-separated series of ellipses. The positions of such distinct substructures are measured in a manner similar to that employed for the HL Tau disk in \citet{2015ApJ...808L...3A}. We perform a preliminary deprojection of each disk image using estimates of the disk orientation from either a Gaussian fit to the image or from measurements of previous ALMA observations of certain axisymmetric sources (\citealt{2018AA...610A..24F} for AS 209, \citealt{2018MNRAS.475.5296D} for Elias 24, and \citealt{2018ApJ...865..157A} for Sz 114 and GW Lup). For each substructure of interest, the radial intensity profile is measured along a series of evenly spaced azimuthal angles in the deprojected image. The azimuthal angle spacing is set to be on the order of a spatial resolution element\textemdash that is, if the FWHM of the major axis of the synthesized beam is $b_\text{maj}$ and the feature occurs roughly at radius $r$ in units of arcseconds (estimated from visual inspection), then the azimuthal angle spacing in degrees is $\Delta \theta \sim 360^\circ \times b_\text{maj}/2\pi r $ (with some minor adjustment so that $\Delta \theta$ is a divisor of 360). In each of the $n$ intensity profile cuts, we search for either a local maximum (if the substructure of interest is a bright ring) or minimum (if the substructure is a gap), generally using a search window of $\sim6$ to 20 au around $r$ depending on the apparent width of the feature. If an extremum cannot be identified in a given intensity profile cut  due to insufficient angular resolution arising from the disk projection or to low SNR, that azimuthal angle is excluded. In addition, the azimuthal angles where the spiral arms cross the gap ($-117^\circ<\theta<-76^\circ$ and $68^\circ<\theta<123^\circ$) are excluded for the Elias 27 disk, and the azimuthal angles spanned by the bright crescent on the southeast side of the HD 143006 disk ($80^\circ<\theta<144^\circ$) are excluded when fitting its outermost ring. The azimuthal angle conventions are explained in Appendix \ref{sec:angles}. 

\begin{figure*}[htp]
\centering
\includegraphics{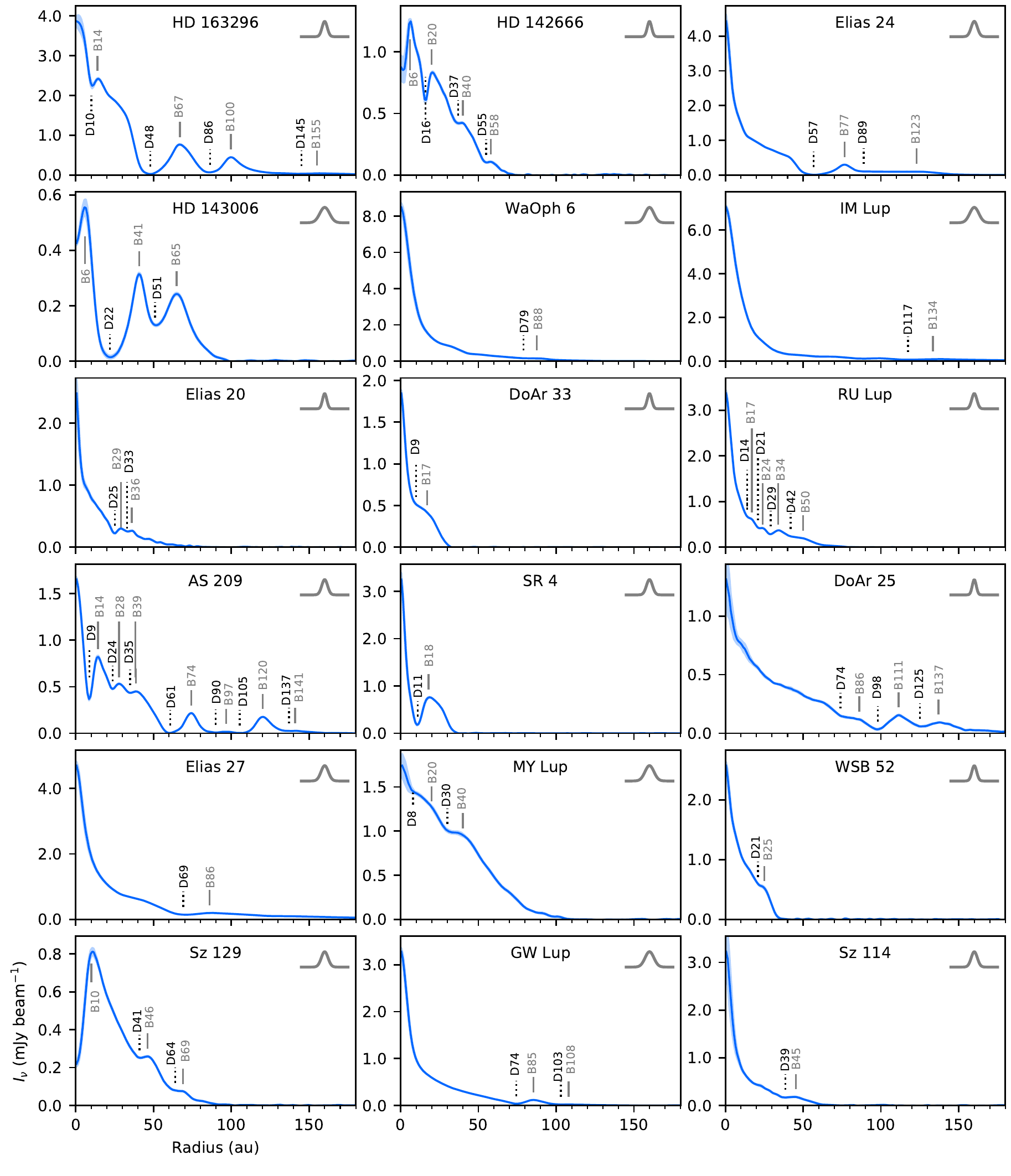}
\caption{Deprojected and azimuthally averaged radial intensity profiles ordered by decreasing stellar luminosity from left to right and top to bottom. Light blue ribbons show the $1\sigma$ scatter at each radial bin divided by the square root of the number of beams spanning the angles over which the intensities are measured. Solid gray lines mark emission rings and dotted black lines mark gaps listed in Table \ref{tab:ringpositions}. The Gaussian profiles show the FWHM of the minor axis of the synthesized beams.\label{fig:linearprofiles}}
\end{figure*}

\begin{figure*}[htp]
\centering
\includegraphics{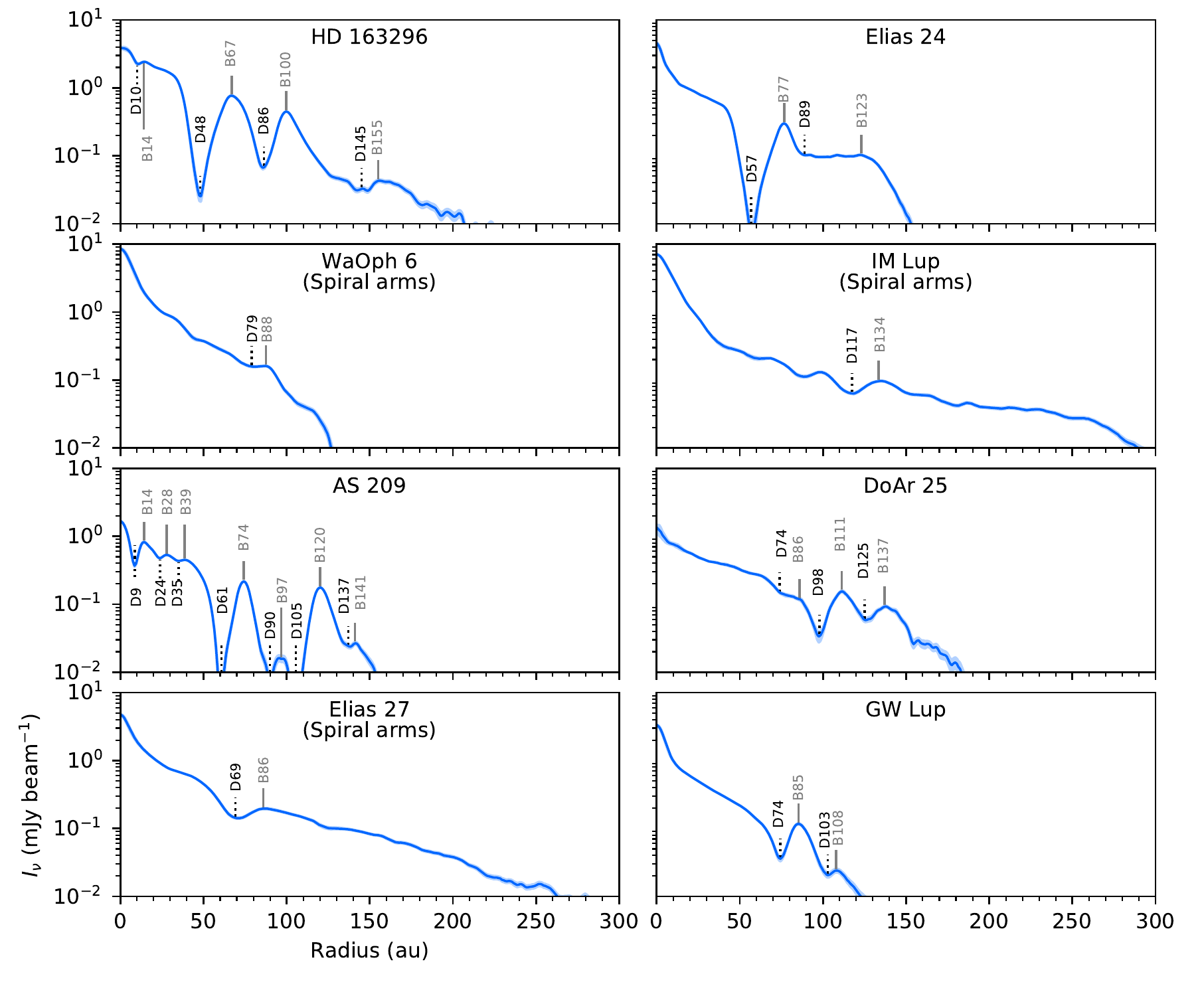}
\caption{Radial intensity profiles plotted on a logarithmic scale for the largest DSHARP disks, ordered by descending stellar luminosity from left to right and top to bottom. Light blue ribbons show the $1\sigma$ scatter at each radial bin divided by the square root of the number of beams spanning the angles over which the intensities are measured. The $y-$axis starts at $10^{-2}$ mJy beam$^{-1}$, corresponding to slightly less than the typical rms level of the continuum images. Note that the Elias 27, IM Lup, and WaOph 6 disks have spiral arms that also create substructures in their radial profiles. \label{fig:logprofiles}}
\end{figure*}

The $n$ locations of the sampled extrema are mapped back to pixel coordinates for the original image. An ellipse is then fit to the $n$ coordinates extracted for a given annular substructure. The model ellipse is defined by five parameters: the offset of the ellipse center from the phase center ($\Delta x$ and $\Delta y$), the semi-major axis of the ellipse ($r_0$), the length ratio of the semi-minor to semi-major axis (i.e., the cosine of the disk inclination, $\cos i$), and the position angle (the angle east of north by which the major axis is rotated). The absolute offsets are not physically significant since phase centers were sometimes adjusted to align execution blocks (see details in \citet{2018ApJ...869L..41A}), but they are necessary for subsequent calculations of the radial profile. The quality of the fit of the model ellipse to the measured coordinates of the annular substructure is assessed by computing the orthogonal distances $d_i$ between the measured coordinates and the model ellipse. The orthogonal distance between a point $(x,y)$ and a curve $f(x,y)$ is the minimum Euclidean distance between the point and the curve. Formulae for calculating orthogonal distances from an ellipse are provided in \citet{zhangconic}.

The log-likelihood takes the form 
\begin{equation}
\ln L = -\frac{1}{2}\sum_{i=1}^n \left[ \frac{d_i^2}{\sigma^2}+\ln(2\pi\sigma^2) \right],
\end{equation}
where $\sigma$ is the standard deviation of the orthogonal distances. It is not straightforward to estimate $\sigma$ directly from the image; although it is tied to the angular resolution, it also depends in some fashion on the SNR of the substructure emission. Thus, in addition to fitting for the five parameters defining the position and orientation of the ellipse, we allow $\ln \sigma^2$ to be a free parameter. Uniform priors are specified for $\Delta x$, $\Delta y$, and $r_0$ over the range $(-4\farcs5, 4\farcs5)$ based on the typical image size. A uniform prior is also specified for $\ln\sigma^2$ over the range $(-15,-5)$, under the assumption that the synthesized beam roughly sets the largest possible value of $\sigma$ and the pixel size roughly sets the lowest. In practice, the best fit values for $\ln \sigma^2$ generally end up being between $-10$ and $-8$, with uncertainties of a few tenths. Gaussian priors for $\cos i$ and position angle (P.A.) are specified for the aforementioned axisymmetric sources with previous ALMA measurements. Otherwise, a uniform prior for $\cos i$ in the range $(0, 1)$ and P.A. in the range (0$^\circ$, 180$^\circ$) are adopted (i.e., the smaller angle east of north is adopted as the position angle, and the direction of the angular momentum vector is not considered because it cannot be determined unambiguously for many of the disks). For HD 143006 and WaOph 6, the bounds for the P.A. are set to (90$^\circ$, 270$^\circ$) because the major axis of the disk appears to be relatively close to 180$^\circ$ \citep[e.g.,][]{2017ApJ...851...83C, 2018AA...619A.171B}.

The posteriors are explored with the affine invariant MCMC sampler implemented in \texttt{emcee} \citep{2010CAMCS...5...65G, 2013PASP..125..306F}. Each ensemble consists of 30 walkers and proceeds for 20,000 steps, with the first 500 steps discarded as burn-in. Convergence of the MCMC chains is assessed by checking that the number of steps significantly exceeds the autocorrelation time (typically on the order of $10^2$). 

The posterior medians for each well-resolved substructure are listed in Table \ref{tab:ringpositions}, with error bars calculated from the 16th and 84th percentiles. If the annular substructure's coordinates were extracted by searching for local minima in the radial profile cuts, it is labeled with the prefix ``D'' (for ``Dark''), followed by the semi-major axis of the best-fit ellipse rounded to the nearest whole number of au. Similarly, if the annular substructure's coordinates were extracted by searching for local maxima, it is labeled with the prefix ``B'' (for ``Bright''), again followed by the semi-major axis of the best-fit ellipse rounded to the nearest whole number of au. This naming convention is adapted from that used for HL Tau in \citet{2015ApJ...808L...3A}. However, instead of naming the substructures based on the order in which they appear in the disk, we use the radial location of the substructure because precise distances are now available from \textit{Gaia} \citep{2016AA...595A...2G, 2018AA...616A...1G}, and we favor a system that can flexibly accommodate additional substructures identified in future observations.

The second part of our procedure addresses features that do not lend themselves well to ellipse-fitting due to lower SNRs or insufficient angular resolution along the projected disk minor axis. The positions of these substructures are instead identified through examination of the deprojected and azimuthally averaged radial intensity profiles. In order to deproject the disks, P.A.s, inclinations and center positions first need to be calculated. For most disks, these quantities are computed by taking weighted averages of the ellipse fits for the annular substructures. For WSB 52, MY Lup, DoAr 33, HD 142666, and SR 4, P.A.s, inclinations, and center positions are instead computed by fitting two-dimensional Gaussians to the images with the \texttt{imfit} task in \texttt{CASA}. Even when an intensity profile deviates from a Gaussian, a Gaussian fit should still provide a good estimate of the source inclination and P.A. provided that the emission is axisymmetric and much larger than the PSF \citep[e.g.,][]{2002AJ....123..583B}. To double-check that Gaussian fitting yields reasonable results, we also fit Gaussians to disks with well-defined rings (and therefore highly non-Gaussian profiles) and find that the results generally agree with those from ellipse-fitting within one degree. The measurements for all disks are listed in Table \ref{tab:orientation}.

Radial intensity profiles are then calculated for each disk by using the previously derived geometries to deproject the disk, binning the pixels in annuli one au wide, and averaging the intensities in each bin. For most disks, the average is taken through all azimuthal angles. For HD 143006 and HD 163296, angles where the outer disk emission asymmetry is visible are excluded from the average ($80^\circ<\theta<144^\circ$ and $50^\circ<\theta<130^\circ$, respectively). For disks with higher inclinations (Elias 20, DoAr 25, MY Lup, WSB 52, HD 142666, and HL Tau), only azimuthal angles that are within $\pm20^\circ$ (as measured in deprojected coordinates) of the projected disk major axis are included in the average because most of the substructures are poorly resolved along the minor axis. The radial profile is calculated in a similar manner for Sz 114 as well due to the extreme elongation of the synthesized beam. Figure \ref{fig:linearprofiles} shows the labeled radial intensity profiles. Figure \ref{fig:logprofiles} provides logarithmically scaled radial intensity profiles of the disks exceeding 100 au in radius in order to show the substructures in the faint outer regions more clearly.

The radial locations of additional annular substructures are identified by searching for local maxima and minima in the averaged radial profiles and cross-referencing with the original image to verify that they correspond to annular features rather than non-axisymmetric features such as spiral arms. The uncertainty on the radial location of each extremum is estimated as the width of the radial bin (i.e., one au). The positions of features measured through ellipse-fitting are checked for consistency with the positions measured from the averaged radial intensity profiles.

Finally, for certain disks (e.g., MY Lup and DoAr 33), there are regions of the radial profile that are ``plateau''-like rather than local maxima or minima. We classify them as annular substructures because they still appear to represent deviations from standard ``smooth'' disk profiles \citep[e.g.,][]{2009ApJ...700.1502A, 2017AA...606A..88T}. This is in part justified by optical depth estimates (Section \ref{sec:opticaldepths}) indicating that many of these ``plateaus'' in the intensity profile may trace local extrema in the surface density profile. The inner and outer edges of the ``plateau'' regions are identified through visual inspection, with some guidance from checking where $\frac{1}{I_\nu(r)}\frac{dI_\nu(r)}{dr}$, the radial intensity profile slope divided by the radial intensity, exceeds $-0.05$ (i.e., where the radial decrease of the intensity is relatively small).  The inner edge of the ``plateau'' is labeled with the prefix ``D'' followed by its radial location in au rounded to the nearest whole number. Similarly, the outer edge of the ``plateau'' is labeled with the prefix ``B'' followed by its radial location. Because of the greater degree of subjectivity involved in identifying the locations of these features compared to identifying local extrema in the emission profiles, approximate locations are listed in Table \ref{tab:ringpositions} without formal error estimates. However, the precision should be on the order of the scale of the synthesized beam (i.e., a few au). 

For comparison, the annular substructure locations for the TW Hya and HL Tau disks are appended to Table \ref{tab:ringpositions}. The small difference in the continuum frequency compared to the DSHARP sources is not expected to significantly affect the  substructure positions. The HL Tau substructure positions from Table 2 of \citet{2015ApJ...808L...3A} are adjusted for the new distance estimate of 147 pc from \citet{2018ApJ...859...33G} and renamed accordingly. The TW Hya substructure positions are measured using the same procedure described for the other sources, except the P.A. is fixed to 155$^\circ$ and the inclination to  7$^\circ$ because nearly face-on disk orientations are more reliably measured from gas observations than from continuum observations \citep{2004ApJ...616L..11Q}. We do not rely on molecular line observations to measure the inclinations of the DSHARP disks because a large fraction of them are cloud-contaminated \citep[see the  $^{12}$CO channel maps in ][]{2018ApJ...869L..41A}. The distance to TW Hya is 60 pc \citep{2016AA...595A...2G}, and radial bins are in increments of $0.5$ au because of the smaller distance. 

Of the DSHARP sample, millimeter continuum substructures have previously been reported for the disks around AS 209 \citep{2017ApJ...835..231H, 2018AA...610A..24F}, DoAr 25 \citep{2017ApJ...851...83C}, Elias 24 \citep{2017ApJ...851L..23C,2017ApJ...851...83C,2018MNRAS.475.5296D}, Elias 27 \citep{2016Sci...353.1519P}, HD 142666 \citep{2018ApJ...860....7R}, HD 143006 \citep{2016ApJ...827..142B}, and HD 163296 \citep{2016ApJ...818L..16Z, 2016PhRvL.117y1101I}.  They have also been  tentatively reported for the disks around IM Lup \citep{2016ApJ...832..110C} and Sz 129 \citep{2017AA...606A..88T}. Although a few studies have classified MY Lup as a transition disk \citep[e.g.,][]{2012ApJ...749...79R,2017AA...606A..88T, 2018ApJ...854..177V}, the emission observed at high angular resolution is still centrally peaked. However, there is radial substructure in the vicinity of where \citet{2017AA...606A..88T} inferred a cavity wall from visibility modeling, which may account for its previous classification as a transition disk. 

The improved resolution of the DSHARP observations reveals additional structural complexity in all disks with previously known or proposed substructures. Millimeter continuum substructures are reported for the first time in the other DSHARP disks. However, the substructure in the WSB 52 disk is only tentatively classified as annular because the inclination of the disk makes it difficult to discern the exact nature of the features observed along the major axis of the projected emission.

\startlongtable
 \begin{deluxetable*}{lllllllllll}
 \tabletypesize{\scriptsize}
\tablecaption{Properties of annular substructures\label{tab:ringpositions}}
\tablehead{
\colhead{Source} &\colhead{Feature} &\colhead{$\Delta x$}& \colhead{$\Delta y$}&\colhead{$r_0$}& \colhead{$r_0$}&\colhead{Incl.}&\colhead{P.A.}&\colhead{Method}&\colhead{Width}&\colhead{Depth}\\
&&\colhead{(mas)}&\colhead{(mas)}&\colhead{(mas)}&\colhead{(au)}&\colhead{(degrees)}&\colhead{(degrees)}&\colhead{}&\colhead{(au)}&\colhead{}}
\colnumbers
\startdata
AS 209 &D9 &$3\pm1$&$-3.9\pm0.9$&$71.8\pm0.8$&$8.69\pm0.11$&$35.6\substack{+0.7 \\ -0.8}$&$85.7\pm0.7$&\text{E}&$4.7\pm0.2$&$0.45\pm0.02$\\
& B14 &$2.0\pm1.3$ &$-4.3\substack{+1.2\\ -1.1}$&$117.6\pm1.1$&$14.2\pm0.14$&$34.9\substack{+0.7\\-0.8}$&$85.8\pm0.7$&E&$8.9\pm0.2$&$-$\\
& D24 &$2.7\pm1.4$ &$-3.8\pm1.2$&$197.0\pm1.3$&$23.84\pm0.15$&$35.6\pm0.7$&$85.8\pm0.7$&E&$3.4\pm0.3^\dagger$&$0.895\pm0.013$\\
& B28&$3\pm2$&$-4\pm2$&$230\pm2$& $27.8\pm0.3$&$35.1\pm0.7$&$86.1\pm0.7$&E&$4.7\pm0.3$&$-$\\
& D35&$4.7\pm1.7$&$-3.5\pm1.5$&$289.6\pm1.6$&$35.04\pm0.18$&$35.2\pm0.6$&$85.4\pm0.7$&E&$3.0\pm0.3^\dagger$&$0.97\pm0.01$\\
& B39 &$4\pm2$&$-3\pm2$&$ 320\pm2$&$38.7\substack{+0.3\\-0.2}$&$34.9\pm0.7$&$85.7\pm0.7$&E&$3.4\pm0.3^\dagger$&$-$\\
& D61&$-2\pm3$&$-4\pm3$&$503\pm3$&$60.8\pm0.3$&$35.8\pm0.6$&$86.1\pm0.6$&E&$15.5\pm0.3$&$0.014\pm0.011$\\
& B74&$-0.8\pm1.1$&$-4.4\pm1.0$&$612.8\pm1.3$ &$ 74.15\pm0.12$&$34.92\pm0.3$ &$ 85.7\pm0.4$&E&$9.3\pm0.3$&$-$\\
& D90&$3\pm3$&$-3\pm3$&$742.6\pm3$&$89.9\pm0.3$&$34.4\pm0.5$&$86.0\pm 0.6$&E&$4.5\pm0.6$&$0.48\pm0.13$\\
& B97&$1\pm5$&$-3\pm4$& $799\pm 4$ & $96.7\pm0.5$ &$ 34.5\pm0.6$ &$ 85.2 \pm0.7$&E&$8.1\pm0.6$&$-$\\
& D105&$1\pm3$&$0\pm3$&$872\pm3$&$105.5\pm0.4$&$35.0\pm0.5$ &$ 85.6 \pm0.6$&E&$14.7\pm0.4$&$0.016\pm0.009$\\ 
& B120 &$0.2\pm1.1$ &$2.1\pm1.0$&$992.9\pm1.3$&$ 120.14\pm0.12$ & $34.92\pm0.19$ & $85.6\pm0.3$&E&$11.2\pm0.4$&$-$\\
& D137&$-$&$-$&$1132\pm8$&$137\pm1$&$-$&$-$&R&$4.2\pm1.4^\dagger$&$0.90\pm0.07$\\
& B141&$-$&$-$&$1165\pm8$&$141\pm1$&$-$&$-$&R&$2.8\pm1.4^\dagger$&$-$\\
\hline
DoAr 25&D74&$-$&$-$&$\sim540$&$\sim74$ &$-$&$-$&V&$-$&$-$\\
&B86&$-$&$-$&$\sim620$&$\sim86$&$-$&$-$&V&$-$&$-$\\
&D98&$40\pm3$&$-495\pm2$&$710\pm3$&$98.0\pm0.3$&$68.2\pm0.3$&$110.9\pm0.3$&\text{E}&$15.5\pm0.5$&$0.22\pm0.04$\\
&B111&$34\pm4$&$-493\pm3$&$806\pm4$&$111.3\pm0.4$&$66.1\substack{+0.3 \\ -0.4}$&$110.2\pm0.4$&\text{E}&$14.3\pm0.5$&$-$\\
&D125&$-$&$-$&$906\pm7$&$125\pm1$&$-$&$-$&R&$10.0\pm1.4$&$0.63\pm0.05$\\
&B137&$-$&$-$&$993\pm7$&$137\pm1$&$-$&$-$&R&$12.8\pm1.4$&$-$\\
\hline
DoAr 33 &D9&$-$&$-$&$\sim60$&$\sim9$&$-$&$-$&V&$-$&$-$\\
 &B17&$-$&$-$&$\sim120$& $\sim17$&$-$&$-$&V&$-$&$-$\\
\hline
Elias 20 & D25 & $-54.4\pm1.5$&$-491.0\pm1.3$&$181.6\pm1.5$&$25.07\pm0.17$&$49\pm1$&$153.2\pm1.3$&E&$3.5\pm1.0$&$0.75\pm0.03$\\
& B29 &$-$&$-$&$210\pm7$&$29\pm1$&$-$&$-$&R&$5.2\pm1.0$&$-$\\
&D33 &$-$&$-$&$239\pm7$&$33\pm1$&$-$&$-$&R&$2.5\pm1.4^\dagger$&$0.95\pm0.03$\\
&B36 &$-$&$-$&$261\pm7$&$36\pm1$&$-$&$-$&R&$1.9\pm1.4^\dagger$&$-$\\
\hline
Elias 24& D57&$110.0\pm1.8$&$-385.8\pm1.9$&$418\pm2$&$56.8\pm0.3$&$27.7\pm0.9$&$51\pm2$&E&$22.8\pm0.3$&$0.03\pm0.01$\\
&B77&$111\pm1$&$-387\pm1$&$564.0\pm1.2$&$76.71\pm0.13$&$29.2\pm0.4$&$45.2\pm0.8$&E&$12.2\pm0.3$&$-$\\
&D89&$-$&$-$&$654\pm7$&$89\pm1$&$-$&$-$&R$$&$-$&$-$\\
&B123\tablenotemark{a}&$-$&$-$&$904\pm7$&$123\pm1$&$-$&$-$&R&$-$&$-$\\
\hline
Elias 27 & D69 &$-5\pm5$&$-8\pm3$&$596\substack{+8\\-7}$&$69.1\pm0.4$&$56.2\pm0.8$&$118.8\pm0.7$&E&$14.3\pm1.1$&$0.73\pm0.02$\\
& B86 &$-$&$-$&$741\pm9$&$86\pm1$&$-$&$-$&R&$21.2\pm1.1$ \\
\hline
GW Lup& D74&$-2.6\pm1.5$&$0.8\pm1.5$&$479.2\pm1.8$&$74.3\pm0.2$&$38.7\substack{+0.4 \\ -0.5}$&$37.7\pm0.8$&E&$12.1\pm0.4$&$0.31\pm0.03$\\
 & B85&$-2.3\pm0.2$&$1.1\substack{+1.8 \\ -1.7}$&$551\pm2$&$85.4\pm0.3$&$38.7\pm0.4$&37.5$\pm0.7$&E&$11.3\pm0.4$&$-$\\
 &D103 &$-$&$-$&$665\pm6$&$103\pm1$&$-$&$-$&R&$4.3\pm1.4^\dagger$&$0.85\pm0.09$\\
 &B108&$-$&$-$&$697\pm6$&$108\pm1$&$-$&$-$&R&$5.5\pm1.4^\dagger$&$-$\\
 \hline
 HD 142666&B6&$-$&$-$&$41\pm7$&$6\pm1$&$-$&$-$&R&$5.3\pm1.4$&$-$\\
 &D16&$-$&$-$&$108\pm7$&$16\pm1$&$-$&$-$&R&$3.5\pm1.4$&$0.73\pm0.02$\\
 &B20&$-$&$-$&$135\pm7$&$20\pm1$&$-$&$-$&R&$7.8\pm1.4$&$-$\\
 &D37&$-$&$-$&$243\pm7$&$37\pm1$&$-$&$-$&R&$<20$&$0.99\pm0.02$\\
 &B40&$-$&$-$&$270\pm7$&$40\pm1$&$-$&$-$&R&$<18$&$-$\\
 &D55&$-$&$-$&$372\pm7$&$55\pm1$&$-$&$-$&R&$2.2\pm1.4^\dagger$&$0.96\pm0.06$\\
 &B58&$-$&$-$&$392\pm7$&$58\pm1$&$-$&$-$&R&$2.1\pm1.4^\dagger$&$-$\\
 \hline
 HD 143006 &B6\tablenotemark{b}&$-$&$-$&$36\pm6$&$6\pm1$&$-$&$-$&R&$5\pm1.4^\dagger$&$-$ \\
 & D22 &$-$&$-$&$133\pm6$&$22\pm1$&$-$&$-$&R&$21.7\pm1.0$&$0.04\pm0.02$\\
 &B41&$-6.3\pm0.6$&$21.5\pm0.6$&$247.3 \substack{+0.8 \\ -0.7}$&$40.8\pm0.1$&$18.9\substack{+0.8 \\ -0.9}$&$169\pm3$&E&$12.2\pm1.0$&$-$\\
 &D51 &$-$&$-$&$309\pm6$&$51\pm1$ &$-$&$-$&R&$12.8\pm1.4$&$0.53\pm0.02$\\
 &B65 &$-1.4\pm1.9$&$24\pm2$&$393\pm3$&$64.9\pm0.4$&$17.2\substack{+1.9 \\ -2.0}$&$164\substack{+7 \\ -6}$&E&$11.5\pm1.4$&$-$\\
 \hline
HD 163296 &D10&$-$&$-$&$100\pm10$&$10\pm1$&$-$&$-$&R&$3.2\pm1.4^\dagger$&$0.93\pm0.03$\\
&B14&$-$&$-$&$140\pm10$&$14\pm1$&$-$&$-$&R&$3.6\pm1.4^\dagger$&$-$\\
&D48&$-$&$-$&$480\pm10$&$48\pm1$&$-$&$-$&R&$20.2\pm1.0$&$0.033\pm0.006$\\
&B67&$-4.3\pm1.2$&$6.6\pm1.2$&$663.8\pm1.6$&$67.04\pm0.12$&$46.9\pm0.2$&$133.4\pm0.3$&E&$15.8\pm1.0$&$-$\\
&D86&$1\pm3$&$2\pm3$&$855\pm4$&$86.4\pm0.3$&$46.9\pm0.4$&$132.7\pm0.5$&E&$16.2\pm0.3$&$0.151\pm0.008$\\
& B100&$-2.4\pm0.9$&$9.0\pm0.9$&$987.0\pm1.2$&$99.69\pm0.09$&$46.63
\pm0.12$&$133.37\pm0.17$&E&$12.0\pm0.3$&$-$\\
&D145\tablenotemark{a}&$-$&$-$&$1450\pm10$&$145\pm1$&$-$&$-$&R&$13.4\pm1.4$&$0.76\pm0.06$\\
&B155&$-$&$-$&$1550\pm10$&$155\pm1$&$-$&$-$&R&$14.6\pm1.4$&$-$\\
\hline
IM Lup &D117&$-3\pm3$&$-2\pm3$&$743\pm4$&$117.4\pm0.5$&$47.5\pm0.5$&$145.2\pm0.7$&E&$15.8\pm0.7$&$0.66\pm0.02$\\
&B134&$0\pm3$&$3\pm3$&$845\pm4$&$133.5\pm0.5$&$47.5\pm0.5$&$143.9\pm0.6$&E&$18.4\pm0.7$&$-$\\
\hline
MY Lup &D8 &$-$&$-$&$\sim50$&$\sim8$&$-$&$-$&V&$-$&$-$\\
&B20 &$-$&$-$&$\sim130$&$\sim20$&$-$&$-$&V&$-$&$-$\\
&D30&$-$&$-$&$\sim190$&$\sim30$&$-$&$-$&V&$-$&$-$\\
&B40&$-$&$-$&$\sim260$&$\sim40$&$-$&$-$&V&$-$&$-$\\
\hline
RU Lup&D14&$-$&$-$&$\sim90$&$\sim14$&$-$&$-$&V&$-$&$-$\\
&B17&$-$&$-$&$\sim110$&$\sim17$&$-$&$-$&V&$-$&$-$\\
&D21&$-$&$-$&$132\pm6$&$21\pm1$&$-$&$-$&R&$<7$&$-$\\
&B24&$-$&$-$&$151\pm6$&$24\pm1$&$-$&$-$&R&$<8$&$-$\\
& D29&$-17.0\pm1.2$&$87.8\substack{+1.2 \\ -1.1}$&$183.0\pm1.5$&$29.1\pm0.2$&$20\pm2$&$117\pm6$&E&$4.5\pm0.3$&$0.784\pm0.013$\\
& B34 & $-17.2\substack{+1.2 \\ -1.1}$&$88.4\pm1.2$&$213.7\pm1.4$&$34.0\pm0.2$&$17\pm2$&$126\pm7$&E&$5.5\pm0.3$&$-$\\
&D42&$-$&$-$&$\sim260$&$\sim42$&$-$&$-$&V&$<16$&$-$\\
&B50&$-$&$-$&$\sim$310&$\sim50$&$-$&$-$&V&$-$&$-$\\
\hline
SR 4 &D11&$-$&$-$&$82\pm7$&$11\pm1$&$-$&$-$&R&$6.3\pm1.4$&$0.23\pm0.02$\\
&B18&$-$&$-$&$134\pm3$&$18\pm1$&$-$&$-$&R&$13.3\pm1.4$&$-$\\
\hline
Sz 114&D39&$0\pm5$&$3\pm4$&$238\pm3$&$38.6\pm0.6$ &$21\pm2$ & $165\pm7$&E&$4.3\pm0.8^\dagger$&$0.94\pm0.05$\\
&B45 &$-2\pm3$&$4\pm3$ & $280\pm3$&$45.4\pm0.5$&$21.6\substack{+1.6 \\ -1.7}$&$165\pm5$&E&$5.4\pm0.8$\\
\hline
Sz 129 & B10 & $-$&$-$&$62\pm6$&$10\pm1$ &$-$&$-$&R&$17.6\pm1.1$&$-$\\
&D41 &$4\pm3$&$4\pm2$&$255\pm3$&$41.0\pm0.4$&$32.4\substack{+1.9\\-2}$&$148\pm4$ &E&$4.1\pm0.6^\dagger$&$0.98\pm0.02$\\
&B46&$7\pm3$&$2\pm2$&$287\pm3$&$46.1\pm0.4$&$35.3\substack{+1.7\\-1.8}$&$153\pm3$&E&$3.5\pm0.6^\dagger$&$-$\\
&D64 &$-$&$-$&$\sim400$&$\sim64$&$-$&$-$&V&$<23$&$-$\\
&B69 &$-$&$-$&$\sim430$&$\sim69$&$-$&$-$&V&$-$&$-$\\
\hline
WaOph 6&D79&$-$&$-$&$642\pm8$&$79\pm1$&$-$&$-$&R&$6.7\pm1.1$&$0.98\pm0.03$\\
& B88&$-244\pm3$&$-361\pm3$&$712\pm4$&$87.5\pm0.4$&$47.3\pm0.7$&$174.2\pm0.8$&E&$3.6\pm1.1^\dagger$\\
\hline
WSB 52 & D21\tablenotemark{c} &$-$&$-$&$\sim150$&$\sim21$&$-$&$-$&V&$-$&$-$\\
&B25\tablenotemark{c}&$-$&$-$&$\sim180$&$\sim25$&$-$&$-$&V&$-$&$-$\\
\hline
\hline
TW Hya\tablenotemark{d}&D1&$-$&$-$&$\sim17$&$\sim1$&$-$&$-$&V&$-$&$-$\\
&B3 &$-$&$-$&$50\pm8$&$3.0\pm0.5$ &$-$&$-$&R&$2.2\pm0.6$&$-$\\
&D26 &$9\pm2$&$13\pm2$&$427.0\pm1.6$&$25.62\pm0.14$&$-$&$-$&E&$4.0\pm0.5$&$0.804\pm0.008$\\
&B30&$-$&$-$&$492\pm8$&$29.5\pm0.5$&$-$&$-$&R&$<6$&$-$\\
&D32&$-$&$-$&$525\pm8$&$31.5\pm0.5$&$-$&$-$&R&$<4$&$0.986\pm0.008$\\
&B33\tablenotemark{a}&$-$&$-$&$550\pm8$&$33\pm0.5$&$-$&$-$&R&$<10$&$-$\\
&D42&$8.2\pm1.5$&$13.7\pm1.5$&$694.0\pm1.1$&$41.64\pm0.09$&$-$&$-$&E&$3.40\pm0.14$&$0.763\pm0.009$\\
&B45&$6.8\pm1.2$&$15.3\pm1.2$&$745.3\pm0.9$&$44.72\pm0.07$&$-$&$-$&E&$2.80\pm0.14$&$-$\\
&D48&$-$&$-$&$\sim800$&$\sim48$&$-$&$-$&V&$<7$&$-$\\
&B52&$-$&$-$&$\sim870$&$\sim52$&$-$&$-$&V&$-$&$-$\\
\hline
HL Tau\tablenotemark{e} &D14 &$-$&$-$&$94.6\pm1.4$&$13.9\pm0.2$&$-$&$-$&$-$&$5.7\pm0.2$&$0.41\pm0.03$\\
&B21&$-$&$-$&$145.6\pm0.7$&$21.4\pm0.1$&$-$&$-$&$-$&$12.9\pm0.2$&$-$\\
&D34&$-$&$-$&$230.6\pm0.7$&$33.9\pm0.1$&$-$&$-$&$-$&$4.9\pm0.1$&$0.67\pm0.02$\\
&B40&$-$&$-$&$272.1\pm0.7$&$40.0\pm0.1$&$-$&$-$&$-$&$6.4\pm0.1$&$-$\\
&D44&$-$&$-$&$299\pm7$&$44\pm1$&$-$&$-$&$-$&$3.2\pm1.4$&$0.92\pm0.02$\\
&B49&$-$&$-$&$333\pm7$&$49\pm1$&$-$&$-$&$-$&$4.2\pm1.4$&$-$\\
&D53&$-$&$-$&$\sim360$&$\sim53$&$-$&$-$&$-$&$<10$&$-$\\
&B58&$-$&$-$&$\sim390$&$\sim58$&$-$&$-$&$-$&$<15$&$-$\\
&D67&$-$&$-$&$458.5\pm0.7$&$67.4\pm0.1$&$-$&$-$&$-$&$4.4\pm0.1$&$0.84\pm0.03$\\
&B72&$-$&$-$&$491.2\pm0.7$&$72.2\pm0.1$&$-$&$-$&$-$&$5.1\pm0.1$&$-$\\
&D77&$-$&$-$&$526.5\pm0.7$&$77.4\pm0.1$&$-$&$-$&$-$&$6.1\pm0.1$&$0.62\pm0.02$\\
&B85&$-$&$-$&$581.0\pm0.7$&$85.4\pm0.1$&$-$&$-$&$-$&$12.0\pm0.1$&$-$\\
&D96&$-$&$-$&$\sim650$&$\sim96$&$-$&$-$&$-$&$<20$&$-$\\
&B102&$-$&$-$&$\sim690$&$\sim102$&$-$&$-$&$-$&$-$&$-$\\
\enddata
\tablecomments{Column descriptions: (1) Name of host star. (2) Substructure label. (3) Right ascension offset from phase center. (4) Declination offset from phase center. (5) Radial location of substructure in mas. (6) Radial location of substructure in au (the uncertainties in mas are simply scaled from the fitting procedure and do not account for the uncertainty in the distance to the source). (7) Inclination of substructure (0$^\circ$ is face-on and 90$^\circ$ is edge-on. (8) Position angle of substructure (east of north). (9) Method used to derive radial location of substructure: ``E'' indicates ellipse-fitting, ``R'' indicates identification of local extrema in the radial profile, and ``V'' indicates identification through visual inspection. (10) Width of substructure (only measured for those identified via ``E'' and ``R'' methods). (11) Depth of gap. All uncertainties are $1\sigma$.}
$^\dagger$ Width of feature is narrower than the minor axis FWHM of the synthesized beam (see Table \ref{tab:orientation} for the synthesized beam size). 
\tablenotetext{a}{This feature shows tentative signs of being multiple gaps and rings.}
\tablenotetext{b}{This ring may be misaligned relative to the outer two rings and therefore could have a larger radius than implied by the radial profile. See \citet{2018ApJ...869L..50P}.}
\tablenotetext{c}{Substructures tentatively classified as annular.}
\tablenotetext{d}{Fit to the 290 GHz continuum image from \citet{2018ApJ...852..122H}. The P.A. and inclination are fixed at 155$^\circ$ and 7$^\circ$, respectively, based on values derived from $^{12}$CO observations in \citet{2004ApJ...616L..11Q}.}
\tablenotetext{e}{Measurements for substructure positions taken from \citet{2015ApJ...808L...3A} and rescaled to a distance of 147 pc \citep{2018ApJ...859...33G}.}
\end{deluxetable*}

 \begin{deluxetable*}{cccccccc}
\tablecaption{Disk geometries\label{tab:orientation}}
\tablehead{
\colhead{Source}&\colhead{$\Delta x$}&\colhead{$\Delta y$} &\colhead{Incl.}&\colhead{P.A.}&\colhead{Method}&\colhead{$R_\text{dust}$}&$\theta_\text{b,min}\times d$\\
&(mas)&(mas)&(deg)&(deg)&&(au)&(au)}
\colnumbers
\startdata
AS 209&$1.9\pm0.5$ &$-2.5\pm0.5$&$34.97\pm0.13$&$85.76\pm0.16$&E&$139\pm1$&4.4\\
DoAr 25&$38\pm2$&$-494\pm2$&$67.4\pm0.2$&$110.6\pm0.2$&E&$165\pm1$&3.0\\
DoAr 33&$1.5\pm0.8$&$0.6\pm0.6$&$41.8\pm0.8$&$81.1\pm1.2$&G&$27\pm1$&3.4\\
Elias 20 &$-54.5\pm1.5$ &$-491.0\pm1.3$&$49\pm1$&$153.2\pm1.3$&E&$64\pm1$&3.2\\
Elias 24 & $110.8\pm0.8$&$-386.8\pm0.9$&$29.0\pm0.3$&$45.7\pm0.7$&E&$136\pm1$&4.7\\
Elias 27 & $-5\pm5$ &$-8\pm3$&$56.2\pm0.8$&$118.8\pm0.7$&E&$254\pm1$&5.5\\
GW Lup & $-2.4\pm1.1$&$0.9\pm1.2$&$38.7\pm0.3$&$37.6\pm0.5$&E&$105\pm1$&6.6\\
HD 142666 &$-43.5\pm0.3$&$28.9\pm0.6$&$62.22\pm0.14$&$162.11\pm0.15$ &G&$59\pm1$&3.3\\
HD 143006 & $-5.9\pm0.6$&$21.7\pm0.6$&$18.6\pm0.8$&$169\pm2$&E&$82\pm1$&7.4\\
HD 163296 &$-2.8\pm0.7$ &$7.7\pm0.7$&$46.7\pm0.1$&$133.33\pm0.15$&E&$169\pm1$&3.9\\
IM Lup &$-1.5\pm2$&$1\pm2$&$47.5\pm0.3$&$144.5\pm0.5$&E&$264\pm1$&6.9\\
MY Lup & $-77.9\pm0.8$&$62.9\pm0.6$&$73.2\pm0.1$&$58.8\pm0.1$&G&$87\pm1$&6.7\\
RU Lup &$-17.1\pm0.8$ &$88.1\pm0.8$&$18.8\pm1.6$&$121\pm5$&E&$63\pm1$&3.9\\
SR 4&$-56.4\pm1.2$&$-507.4\pm1.3$&$22\pm2$ &$18\pm5$&G&$31\pm1$&4.6\\
Sz 114 & $-1\pm2$&$4\pm2$&$21.3\pm1.3$&$165\pm4$&E&$58\pm1$&4.6\\
Sz 129 & $5.4\pm1.9$&$3.0\pm1.7$&$34.1\pm1.3$&$151\pm2$&E&$76\pm1$&5.0\\
WaOph 6 &$-244\pm3$ &$-361\pm3$&$47.3\pm0.7$&$174.2\pm0.8$&E&$103\pm1$&6.6\\
WSB 52 &$-119.5\pm0.4$&$-432.8\pm0.4$&$54.4\pm0.3$& $138.4\pm0.3$&G&$32\pm1$&3.7\\
\enddata
\tablecomments{Column descriptions: (1) Right ascension offset from phase center of disk. (2) Declination offset from phase center of disk. (3) Disk inclination. (4) Disk position angle. (5) Method used to derive disk P.A., inclination, and phase offset. ``E'' denotes that the offset from phase center, P.A., and inclination are derived by fitting ellipses to individual annular substructures. ``G'' denotes that these quantities are estimated by fitting a 2D Gaussian to the image. (6) Radial extent of 1.25 mm continuum emission. (7) FWHM of the synthesized beam along the minor axis multiplied by the distance to the source (see Table 4 of \citet{2018ApJ...869L..41A} for beam dimensions, distance to source, and references).}
\end{deluxetable*}

\subsection{Widths and contrasts of annular substructures\label{sec:widths}}

Defining the sizes of disk substructures is a more nebulous task compared to measuring their radial positions. The widths and amplitudes of emission rings are sometimes measured by modeling them as Gaussian rings, as is done for the DSHARP observations of AS 209, GW Lup, Elias 24, HD 143006, and HD 163296 \citep{2018ApJ...869L..46D,2018ApJ...869L..48G,2018ApJ...869L..49I,2018ApJ...869L..50P}. These sources lend themselves well to such a procedure because of the high contrast and large separation of many of their substructures. In other cases where the Gaussian components overlap substantially or where the substructures do not closely resemble Gaussians, it is less useful to define ring widths in terms of Gaussian parameters. In disk simulations, the widths and depths of gaps are often defined with respect to an initial surface density profile \citep[e.g.,][]{2015ApJ...807L..11D, 2016PASJ...68...43K}, but these definitions do not translate well to observations, where the initial surface density profile is unknown. 

Instead, similarly to \citet{2018ApJ...869L..47Z}, we take a more empirical approach and define width and depth without assuming a functional form for the substructures or an initial surface density. For an adjacent gap-ring pair, we define the gap radius $r_d$ as the value of $r_0$ listed in Table \ref{tab:ringpositions} for a substructure labeled with the prefix ``D'' and the reference ring radius $r_b$ as the $r_0$ value listed for the ``B''-labeled substructure directly outside the gap. The gap depth is then defined as 
\begin{equation}
\frac{I_d}{I_b},
\end{equation}
where the gap intensity $I_d$ is the mean intensity value in the radial bin containing $r_d$ and the reference intensity $I_b$ is the mean intensity value in the radial bin containing $r_b$. The uncertainty adopted for the mean intensity value is $\sfrac{\sigma_\text{bin}}{\sqrt{N_\text{beam}}}$, where $\sigma_\text{bin}$ is the $1\sigma$ scatter of the pixel values in the radial bin and $N_\text{beam}$ is the number of synthesized beams spanned by the ellipse or arc over which the intensity is being averaged. 

For measurements of substructure widths, the radial intensity profiles are then linearly interpolated onto radial gridpoints spaced by 0.1 au. The dividing point between the outer edge of a gap and the inner edge of its immediate exterior ring is defined as the radius $r_{d,o}$ (where $r_d<r_{d,o}<r_b$)  at which the intensity is equal to $I_\text{mean} = 0.5(I_d+I_b)$. The inner edge of the gap $r_{d,i}$ is defined as the largest value $r$ satisfying the criteria that $I(r) = I_\text{mean}$ and $r<r_d$. The outer edge of a ring $r_{b,o}$ is defined as the smallest value $r$ satisfying the criteria that $I(r) = I_\text{mean}$ and $r>r_b$. Thus, the width of a gap is $r_{d,o}-r_{d,i}$ and the width of a ring is $r_{b,o}-r_{d,o}$. For rings with emission profiles that can be modeled as isolated Gaussians \citep[e.g.,][]{2018ApJ...869L..46D}, this definition of ring width is essentially the FWHM. A diagram is shown in Appendix \ref{sec:schematic}, and a couple special cases are addressed. The measurements are listed in Table \ref{tab:ringpositions}. These definitions of width and depth are only applicable to the ``D''/``B'' pairs that correspond to local extrema in the radial profiles, i.e., those with measurement methods labeled either ``E'' or ``R.''  Features can have measured widths smaller than the synthesized beam because the peak-to-peak or trough-to-trough separations of the gaps/rings on either side of the feature are still larger than the beam.

For shallow emission features that occur within a series of annular substructures, upper bounds for the widths can be estimated by measuring the peak-to-peak or trough-to-trough distances of the adjacent substructures.  It is not clear whether the shallowest emission features correspond to surface density features that are shallow and wide or deep and narrow. Observations at higher sensitivity and angular resolution are necessary to measure the widths and contrasts of shallow substructures. 

These definitions have a few important limitations. First, the effects of the beam are not accounted for. Thus, the true gap widths should generally be larger and the ring widths should be narrower. In addition, the true ratio of $I_d$ to $I_b$ should generally be smaller because beam smearing will reduce peak intensities and ``fill in'' gaps. For well-resolved features, the effect of beam convolution on the width and depth estimates should not be too significant, but the effects are quite significant for features near the angular resolution limit. Therefore, the values quoted for such narrow features may be better viewed as upper or lower bounds, as appropriate. Second, the gap contrast is defined with respect to the exterior ring because a local maximum is guaranteed to occur outside a local minimum in the radial intensity profile, whereas there is not necessarily a ring interior to a gap. However, this can sometimes lead to counterintuitive outcomes, such as  D90 in the AS 209 disk having a relatively shallow measurement for gap depth even though the gap intensity is at most a few percent of the peak intensity of the immediately interior B74 feature. Similarly, if the local minimum outside a ring has a much lower intensity value than the local minimum interior to a ring, such as B39 in the AS 209 disk or B65 in the HD 143006 disk, it is not clear whether it is more sensible to view the ring as a narrow enhancement on top of a background profile (which our definition implicitly does) or as a single wide structure without additional background emission (as is done when disk emission is modeled as a series of Gaussian rings). Overall, though, the definitions we adopt are still useful for comparing substructures within the sample and for assessing whether the features are resolved.

\begin{figure*}[htp]
\centering
\includegraphics{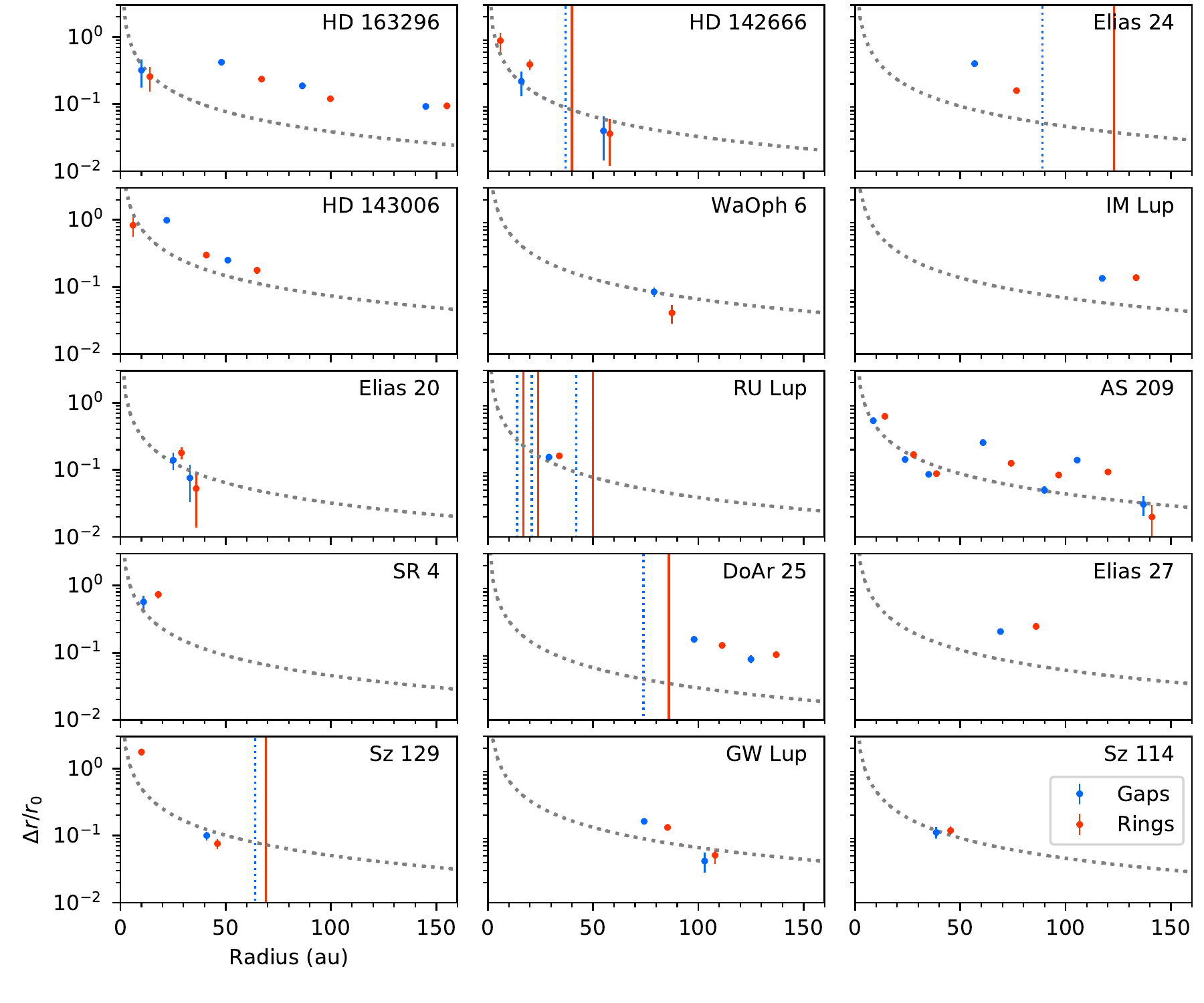}
\caption{Plots of substructure widths divided by radial positions for the 15 disks with sufficiently high-contrast substructures. The positions of substructures that are too shallow to allow a width measurement are marked with blue (gap) or orange (ring) vertical lines. The gray dotted curves correspond to $\frac{\theta_\text{b,min}\times d}{r}$, separating resolved features above the curve from marginally resolved or unresolved features at or below the curve. \label{fig:widths}}
\end{figure*}

The substructure widths and contrasts span a wide range. Substructures range from widths of a few au to more than 20 au, but the majority of the substructures are narrower than 10 au. Variations can be quite wide even within the same sources, as exemplified by AS 209 and HD 163296. Neither substructure widths nor depths in the DSHARP sources increase monotonically with distance from the central star.  A majority of the disks have substructures that are marginally resolved (not necessarily exclusively so), suggesting that there could be additional substructures under the resolution limit. The deepest gaps have intensity levels that are at most a few percent of the peak intensity of adjacent rings, but more typically the intensity variations between adjacent gaps and rings are less than $20\%$. However, many of the low-contrast gaps have widths comparable to the size of the synthesized beam, so their apparent shallowness may result from insufficient resolution.

Figure \ref{fig:widths} shows the ratio of substructure widths to their radial positions, $\sfrac{\Delta r}{r_0}$, for all but the lowest contrast disks (i.e., MY Lup, DoAr 33, and WSB 52). The measured values are generally largest in the inner 20 au of these disks, which is a limitation imposed by angular resolution, as shown by the gray curves in Figure \ref{fig:widths}. Otherwise, values often fall between 0.05 and 0.2, comparable to the typical values for disk aspect ratios, $\sfrac{h}{r}$ \citep[e.g.,][]{2010ApJ...723.1241A}. Equivalently, the widths are often comparable to expected values for the pressure scale height. A few gaps in the outer disk have notably high values of $\sfrac{\Delta r}{r_0}>0.25$: D48 in the HD 163296 disk, D57 in the Elias 24 disk, D22 in the HD 143006 disk, and D61 in the AS 209 disk. These features are among the widest DSHARP gaps in absolute, not simply relative, terms. They are also among the deepest gaps observed, with intensities inside the gap being at most a few percent of those of adjacent rings.

\begin{figure*}[htp]
\centering
\includegraphics{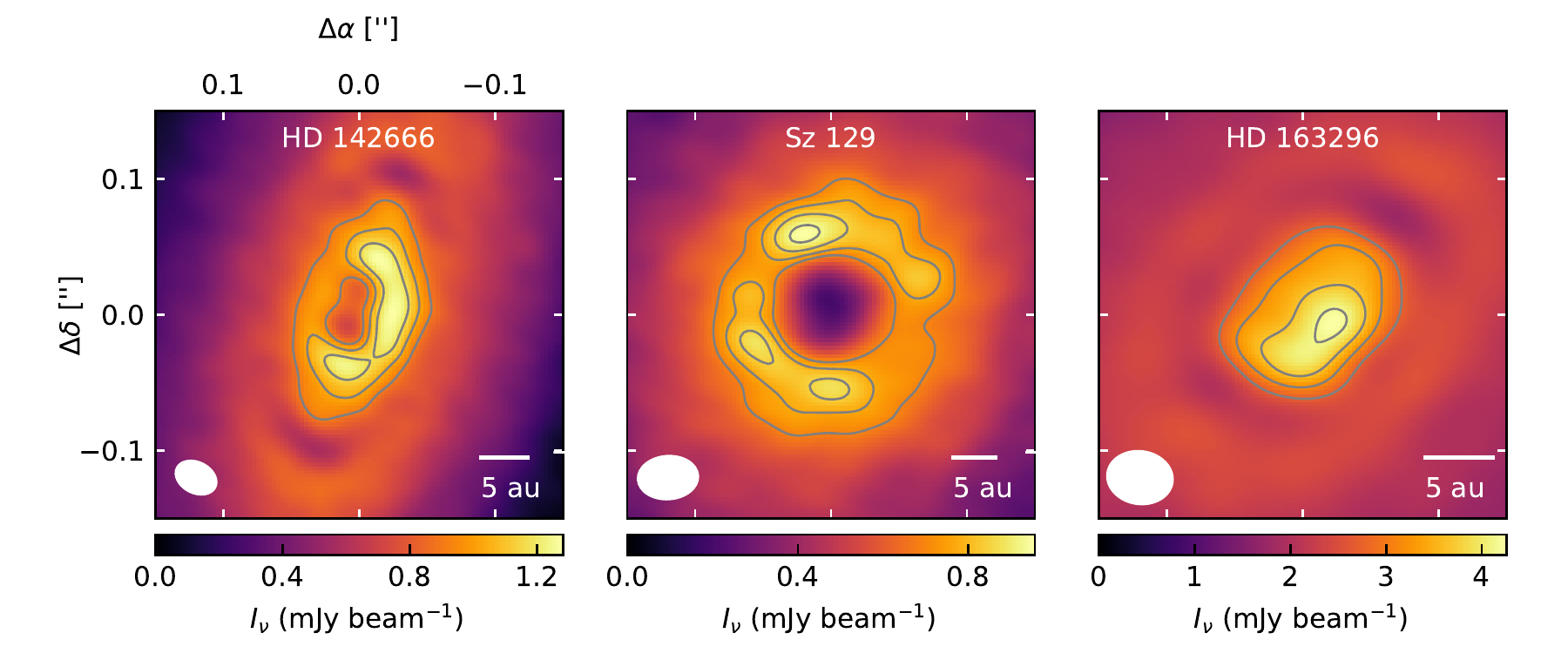}
\caption{Insets of the inner disk brightness asymmetries of HD 142666, Sz 129, and HD 163296. Gray contours are drawn at $[70,80,90]\times$ rms for HD 142666, $[45,50,55,60]\times$ rms for Sz 129, and $[120,140,160,180]\times$ rms for HD 163296. \label{fig:innerdisks}}. 
\end{figure*}

\subsection{Dust disk radii}
To assess how substructure characteristics relate to disk size, we also compute the radius of the millimeter disk, $R_\text{dust}$, for each source using a ``curve-of-growth''-type method similar to that described in  \citet{2016ApJ...828...46A}. For each disk, continuum fluxes are measured within a series of increasingly large elliptical apertures with the same P.A. and major axis/minor axis ratio as the projected disk image. The total continuum flux is taken to be the asymptotic value of the flux curve. $R_\text{dust}$ is defined to be the radius at which the enclosed flux is equal to 95$\%$ of the total flux. The values are listed in Table \ref{tab:orientation}. The uncertainty is estimated as the spacing between elliptical apertures (1 au, limited by the pixel size); the relative uncertainty in the continuum flux does not contribute substantially to the uncertainty in $R_\text{dust}$ due to the high SNR of the images.

\subsection{Additional substructures\label{sec:additionalstructures}}
Sz 129, HD 143006, and HD 142666 exhibit decreases in intensity toward the center of the disk. The angular resolution is not sufficient to determine whether these features are cavities or gaps encircling inner disks. However, high resolution millimeter continuum observations that have detected inner disks inside the cavities of a few transition disks suggest that the distinction between a cavity and an annular gap could largely be a matter of angular resolution \citep[e.g.,][]{2016ApJ...820L..40A, 2017ApJ...840...60B, 2018ApJ...860..124D}. 

The innermost emission ring of Sz 129 is brighter in the northeast and that of HD 142666 is brighter on the southwest side. The millimeter continuum intensity peak of the HD 163296 disk is also offset southwest from the estimated center of the disk by $\sim25$ mas and may similarly be tracing an inner emission ring. Insets of these disks are shown in Figure \ref{fig:innerdisks}. For both HD 163296 and HD 142666, the $^{12}$CO $J=2-1$ emission indicates that the continuum brightness asymmetries in the inner disk are on the far side of the disk \citep{2018ApJ...869L..41A, 2018ApJ...869L..49I}. Thus, the brightness asymmetry may arise from viewing the heated and puffed up interior of a ring. The absolute geometry of Sz 129 is not immediately clear from the $^{12}$CO observations.

A few disks have tentative annular substructures in addition to what is listed in Table \ref{tab:ringpositions}. The radial profile of DoAr 25 shows some departures from smoothness inward of D76. However, since the PSF for this disk is poorer than most of the other DSHARP disks due to the abbreviated integration time, better $uv$ coverage will be needed to confirm the structure of the inner disk. The radial profile of Elias 24 has hints of very low amplitude variations between D89 and B123. The emission profile in this region is also unusual among the DSHARP disks for overall being quite flat over a radial span of tens of au, as best seen in Figure \ref{fig:logprofiles}.  The radial profile of Elias 20 shows some low-amplitude variations outside B36, but additional annular substructures are not obvious in the image (note that this is one of the sources for which the radial profile is only averaged through wedges around the major axis). The disks with spiral arms (WaOph 6, IM Lup, and Elias 27) may have additional gaps and rings beyond what is listed in Table \ref{tab:ringpositions}, but as discussed further in \citet{2018ApJ...869L..43H}, these features may instead be branches or tightly wrapped extensions of the spiral arms. 

Finally, HD 163296 has a crescent-like asymmetry interior to B67 and HD 143006 has an asymmetry exterior to B65. These are discussed further in \citet{2018ApJ...869L..49I} and \citet{2018ApJ...869L..50P}, respectively.

\begin{figure*}[htp]
\centering
\includegraphics{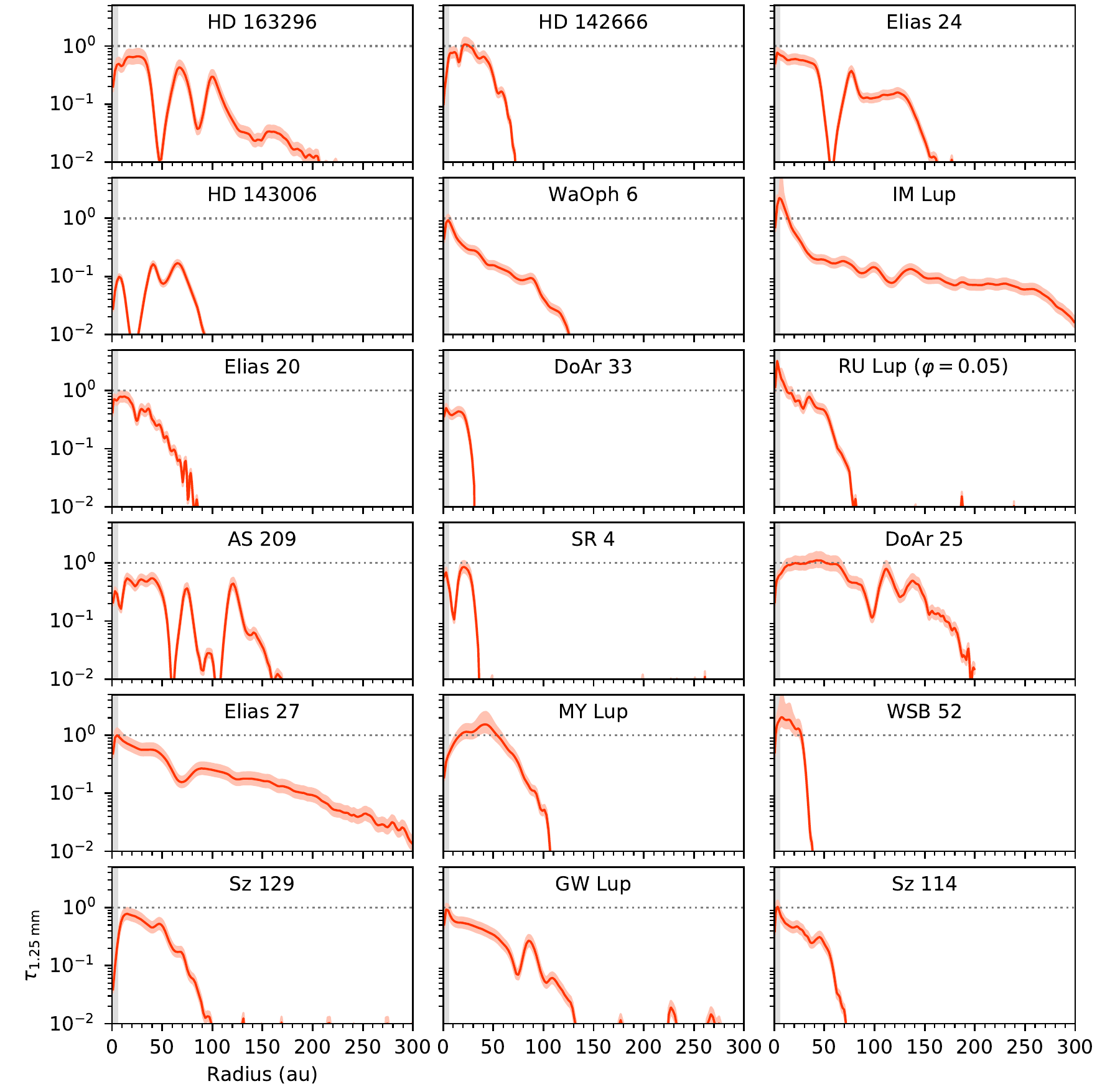}
\caption{Estimated 1.25 mm continuum optical depths profiles for the DSHARP sample, ordered by descending stellar luminosity from left to right and top to bottom. The light orange ribbon shows the range of $\tau$ values using the $1\sigma$ errors for the stellar luminosity from \citet{2018ApJ...869L..41A}. The inner 5 au of each profile is shaded gray to show where beam dilution artificially lowers the derived optical depth. The gray horizontal dotted lines show where $\tau=1$. The $y-$axis starts at $10^{-2}$, commensurate with the optical depth value that the DSHARP observations are expected to be sensitive to. For the disks with radial profiles averaged through narrow azimuthal wedges (i.e. HD 142666, Elias 20, DoAr 25, MY Lup, WSB 52, and Sz 114), the optical depth profiles are only plotted out to where the radial intensity profile drops to $\sim2\times$ the rms, since the noise fluctuations in the profile are larger than for the disks averaged through all azimuthal angles.\label{fig:opticaldepths}}
\end{figure*}

\subsection{Continuum optical depths \label{sec:opticaldepths}}
It has usually been assumed that the millimeter dust continuum of disks is optically thin \citep[e.g.,][]{1990AJ.....99..924B}. Recent analyses of disk spectral indices and emission sizes have explored whether disks only appear to be optically thin at low resolution because of beam dilution of optically thick substructures \citep[e.g.,][]{2010AA...512A..15R, 2017ApJ...845...44T}, motivating an analysis of whether the DSHARP substructures are optically thick. Multiwavelength observations are desirable for constraining dust optical depths \citep[e.g.,][]{2012ApJ...760L..17P,2016AA...588A..53T}, but a rough estimate can still be obtained from the single-wavelength DSHARP observations by approximating the midplane temperature profile with the simplified expression for a passively heated, flared disk in radiative equilibrium \citep[e.g.,][]{1997ApJ...490..368C, 1998ApJ...500..411D,2001ApJ...560..957D}: 
\begin{equation} \label{eq:temperature}
T_\text{mid}(r) = \left( \frac{ \varphi L_\ast}{8\pi r^2 \sigma_\text{SB}}\right)^{0.25},
\end{equation}
where $\sigma_\text{SB}$ is the Stefan-Boltzmann constant, $L_\ast$ is the stellar luminosity (taken from Table 1 of \citet{2018ApJ...869L..41A}), and $\varphi$ is the flaring angle. The optical depth $\tau_\nu$ is then calculated using the relationship 
\begin{equation}
I_\nu (r) = B_\nu(T_\text{mid}(r))(1-\exp(-\tau_\nu (r))), 
\end{equation}
where $I_\nu(r)$ is the deprojected, azimuthally averaged radial intensity profile. The full Planck expression is used because the Rayleigh-Jeans approximation loses accuracy at millimeter wavelengths. 

Since colder temperature profiles lead to a higher estimate of $\tau$ and we wish to examine how large the optical depths can possibly be, we select a somewhat conservative value of $\varphi=0.02$. This corresponds to $\sfrac{h}{r} \approx 0.07 $ at $r=100$ au for a star of solar mass and luminosity. The value of $\varphi$ is also chosen to be consistent with the analysis of a subset of the DSHARP sources in \citet{2018ApJ...869L..46D}, which contains a more detailed discussion of the effect of the assumed temperature profile on other derived disk properties. A higher value of $\varphi=0.05$ is selected for the RU Lup disk because lower values yield dust temperature estimates below the brightness temperature of the inner disk.  Assuming typical values for the beam size ($\sim50$ mas), rms ($1.5\times10^{-5}$ Jy beam$^{-1}$), and disk temperature (10 K), the DSHARP observations are sensitive to dust optical depths on the order of $10^{-2}$. Given uncertainties in $\varphi$ and $L_\star$, the optical depth estimates may be up to a factor of two off from the true optical depths (there is also a $10\%$ absolute flux calibration uncertainty, but this is swamped by the uncertainties in the thermal profile). 

The 1.25 mm continuum optical depth profiles of the DSHARP sample are shown in Figure \ref{fig:opticaldepths}. All of the profiles exhibit an apparent decrease in optical depth in the inner few au, which is likely primarily due to beam dilution. The central dips in the profiles for HD 142666, Sz 129, and HD 143006 are also in part due to bona fide optical depth decreases corresponding to the small emission cavities detected. The MY Lup and DoAr 25 optical depth profiles have relatively broad dips toward the disk center, but the extent to which the highly inclined viewing geometry contributes to this is not clear. At high inclinations, deprojection becomes less reliable due to disk vertical structure, so multidimensional radiative transfer modeling would be more appropriate to estimate the optical depths of these two disks. 

Factoring in the uncertainty in the stellar host luminosity, Figure \ref{fig:opticaldepths} indicates that the millimeter continuum optical depths of most of the DSHARP disks are plausibly of order unity out to radii of 5 to 10 au. This is consistent with the finding from \citet{2018ApJ...859...21A} that most of the disks observed in the Lupus star-forming region have cores that are optically thick at millimeter wavelengths. IM Lup, RU Lup, and WSB 52 appear to have quite substantial optically thick cores that extend to radii of 15 au or beyond. \citet{2016ApJ...832..110C} previously concluded that the inner disk of IM Lup was optically thick at submillimeter wavelengths based on radiative transfer modeling of the 870 $\mu$m continuum and the SED. The low continuum optical depth of the HD 143006 rings (aside from the asymmetric feature on B65 that is $\sim3x$ brighter than the rest of the ring) makes it unusual among the DSHARP sources. Otherwise, the estimated optical depths at the bright rings in the DSHARP sources typically range from 0.3 to 0.6. Observations of decrements in $^{12}$CO $J=2-1$ emission at the locations of the outer bright continuum emission rings for the AS 209 and HD 163296 disks suggest that these substructures may be optically thick despite the moderate estimates shown in Figure \ref{fig:opticaldepths} \citep{2018ApJ...869L..49I, 2018ApJ...869L..48G}. \citet{2018ApJ...869L..46D} propose three possible reasons for this discrepancy: the temperatures are overestimated, scattering is significant, or the dust rings consist of optically thick clumps that only appear to be optically thin due to beam dilution. 

While the absolute values of the optical depth profiles are highly uncertain, the relative values within each disk are better-constrained because they only depend on the power-law exponent of the temperature profile. Outside the regions where the continuum is optically thick, the optical depth profile roughly traces the product of the dust surface density $\Sigma_d$ and the dust opacity $\kappa_\nu$. In the disks around HD 163296, Elias 24, HD 143006, AS 209, SR 4, and DoAr 25, local variations in the radial optical depth profiles reach one to two orders of magnitude, indicating extremely large local variations in the surface density and/or the dust opacity. Large variations in dust opacity would suggest dramatic changes in grain properties across the gap. To verify this, multiwavelength observations to measure the spectral index would be necessary.

\begin{figure}[htp]
\centering
\includegraphics{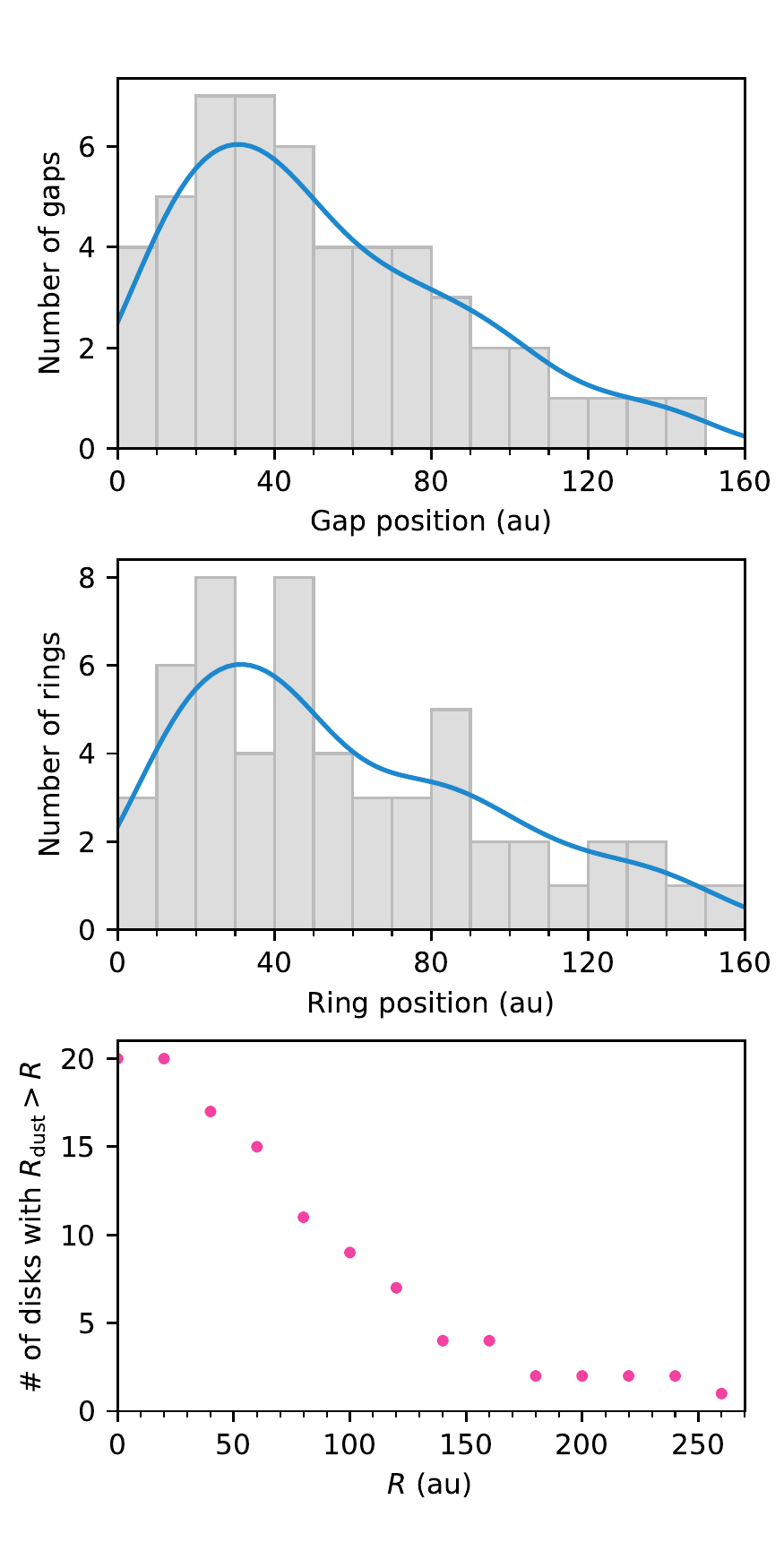}
\caption{\textit{Top:} A histogram of the radial positions of the gaps in the DSHARP sample and the HL Tau and TW Hya disks. The Gaussian kernel density estimate (scaled to match the histogram area) is shown in blue. \textit{Center:} Same as above, but for the bright rings. \textit{Bottom:} Plot of the number of disks where $R_\text{dust}$ exceeds radius $R$.  \label{fig:positionhistogram}}
\end{figure}

\section{Aggregate characteristics of annular substructures}\label{sec:trends}
\subsection{The radial distribution of annular substructures}

Histograms of the positions of the radial substructures of the DSHARP sample, along with HL Tau and TW Hya, are shown in Figure \ref{fig:positionhistogram}. We also use the \texttt{scikit-learn} implementation of Gaussian kernel density estimation to estimate the distribution. The bandwidth is optimized through leave-one-out cross-validation to ensure that it is narrow enough to capture significant features but wide enough to smooth over spurious features. The resulting estimates are plotted over the histograms.

Annular substructures occur at virtually all radii probed by millimeter emission, as far inward as $r<5$ au (i.e., essentially within ALMA's resolution limits) and as far out as $\sim155$ au. The dropoff in the number of substructures detected beyond a radius of $\sim40$ au appears to be a limitation imposed by the overall radial extent of the disk. The bottom of Figure \ref{fig:positionhistogram} shows a ``survival function'' for the disk size, plotting the number of disks with $R_\text{dust}>R$ over a range of values of $R$. Among the DSHARP disks, it is quite common for annular substructures to occur near the outer edge of the detected millimeter continuum emission. Elias 27 and IM Lup are notable exceptions, with their millimeter continuum emission extending more than 100 au beyond the farthest detected annular substructures. The presence of spiral arms and the faintness of emission in the outer disk, though, may complicate the identification of more distant substructures. Substructures can occur at even larger radii than what is measured for the DSHARP sample\textemdash the V1094 Sco disk has an emission ring located at $r\sim220$ au \citep{2018AA...616A..88V}.

Intuitively, one might expect the distribution of substructures to decrease monotonically with radius because the disks have varying sizes and the more compact disks by definition cannot have substructures at large radii. Thus, the slight dropoff in the occurrence of substructures at smaller radii is also interesting to consider. Selection effects must play at least a partial role, since the DSHARP sample excluded known transition disks, which by definition have substructure in the inner disk. 

The optical depth profiles in Figure \ref{fig:opticaldepths} suggest that surface density substructures may be obscured in the intensity profile due to the steep temperature changes in the inner disk. Some disks, such as DoAr 33 and WSB 52, have substructures that appear to be ``plateaus'' in the radial intensity profiles but correspond to distinct local maxima and minima in the radial optical depth profiles. In addition, although the inner 30 au of the Elias 24 and GW Lup images look featureless, their optical depth profiles show slight dips suggestive of additional substructure. However, more accurate temperature profiles derived through radiative transfer modeling are needed to investigate this possibility further.  

Substructures may also be more difficult to detect in the inner disk due to smaller characteristic scales and higher optical depths. The scale height, which increases with distance from the star, influences the width of gaps opened by zonal flows and planets \citep[e.g.,][]{1999ApJ...514..344B, 2009ApJ...697.1269J}. Even if surface density substructures in the inner disk are sufficiently wide, they may still be obscured in the inner disk due to high optical depth. As shown in Figure \ref{fig:opticaldepths}, multiple disks appear to have 1.25 mm continuum optical depths of order unity inside radii of 10 au. Notably, the WSB 52 disk appears to be optically thick throughout much of its extent and has the most marginal substructure detection. 

 \begin{deluxetable}{ccccc}
\tablecaption{Major icelines in disk regions probed by ALMA\label{tab:icelines}}
\colnumbers
\tablehead{
\colhead{Species} &\colhead{Abundance}&\colhead{$E_b$}&Reference&\colhead{$T_\text{frz}$}\\
&&(K)&&(K)}
\startdata
CO$_2$ & 3e$-5$\tablenotemark{a}&2605\tablenotemark{d}&\citealt{2014AA...564A...8M}&$57-72$\\
CO & 5e$-5$&1155\tablenotemark{c}&\citealt{2016ApJ...816L..28F}&$26-32$\\
    & 5e$-5$&866\tablenotemark{d} & \citealt{2016ApJ...816L..28F}&$19-24$\\
N$_2$&4e$-5$\tablenotemark{b}&770\tablenotemark{d}&\citealt{2016ApJ...816L..28F}&$17-21$\\
\enddata
\tablecomments{(1) Species. (2) Abundance with respect to total number of H atoms. (3) Binding energy.  (4) Reference for $E_b$ value. (5) Freezeout temperatures computed for gas number densities from $10^8-10^{12}$ cm$^{-3}$ (chosen to be representative of midplane values).}
\tablenotetext{a}{Following \citet{2011ApJ...743L..16O}, the CO$_2$ abundance is estimated from ice measurements toward the CBRR 2422.8-3423 disk from \citet{2006AA...453L..47P}}
\tablenotetext{b}{Following \citet{2016ApJ...833..203P}, we adopt the proto-Sun elemental nitrogen abundance of $8\times10^{-5}$ with respect to elemental hydrogen \citep{2003ApJ...591.1220L} and assume that N$_2$ is the dominant nitrogen carrier.}
\tablenotetext{c}{Compact H$_2$O substrate.}
\tablenotetext{d}{Pure ice substrate.}
\end{deluxetable}

\subsection{Relationship with temperature\label{sec:temperatures}}
\begin{figure*}[htp]
\centering
\includegraphics{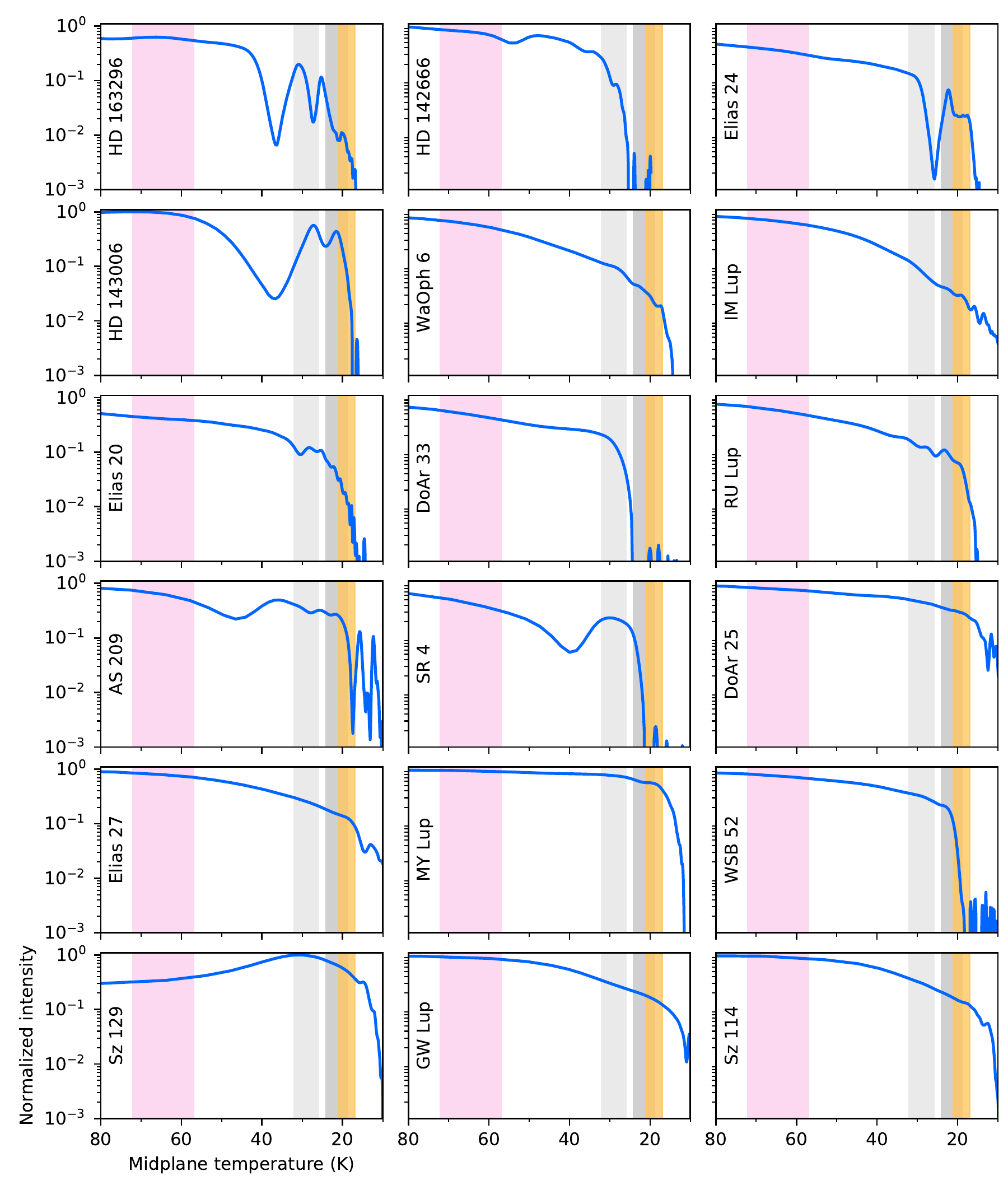}
\caption{ Normalized, azimuthally averaged intensity profiles (in descending order by stellar luminosity) as a function of midplane temperature. Shaded regions show the freezeout temperature ranges for CO$_2$ (pink), CO on water ice (light gray), CO on pure ice (dark gray), and N$_2$ on pure ice (orange). The plotted ranges for CO and N$_2$ on pure ice overlap due to the range of gas densities explored, but the CO snowline will be interior to the N$_2$ snowline in a given disk.\label{fig:snowlines}}
\end{figure*}

Using the radial temperature profiles calculated in Section \ref{sec:opticaldepths}, Figure \ref{fig:snowlines} shows the normalized, azimuthally averaged intensity profiles as a function of midplane temperature from 80 to 10 K. This temperature range is chosen to include the freezeout temperatures of key disk volatiles, since several works have proposed that annular substructures trace the locations of various molecular snowlines \citep[e.g.,][]{2015ApJ...806L...7Z, 2016ApJ...821...82O,2017ApJ...845...68P}. To examine whether the DSHARP substructures are connected to snowlines, we shade in the freezeout temperature ranges for CO$_2$, N$_2$, and CO,  following the approach in \citet{2009ApJ...690.1497H} and using the binding energies and abundances listed in Table 3. While H$_2$O, NH$_3$, and CH$_3$OH are other volatiles expected to be abundant in disks based on cometary measurements and ice observations toward star-forming regions \citep[e.g.,][]{2011ARAA..49..471M,2011ApJ...740..109O}, their binding energies are so high that their snowline locations are expected to be below the resolution limit of the DSHARP observations for all but a couple of the most luminous sources (it is worth noting that HD 163296 and HD 142666, the sources with the highest stellar luminosities, have substructures in their inner 5 au).

There does not appear to be a strong correspondence between temperature and substructure positions. More generally, there is no obvious relationship between stellar luminosity (which controls the temperature estimate) and the substructure positions. With respect to specific snowlines, few substructures are observed within the expected vicinity of the CO$_2$ snowline. About half the disks have substructures that occur in the expected vicinity of the CO and N$_2$ snowlines. However, the possible range of locations for the CO snowline can be quite wide depending on whether CO is largely binding to water ice or pure ice. Moreover, there is a high level of dissimilarity in substructures that occur near the expected locations of the CO and N$_2$ snowlines; some disks have a series of narrow annuli while others have a single wide, deep gap. Thus, it seems unlikely that all of these substructures are directly associated with the CO and N$_2$ snowlines.

\subsection{The spacings of annular substructures\label{sec:spacings}}
\begin{figure}[htp]
\centering
\includegraphics{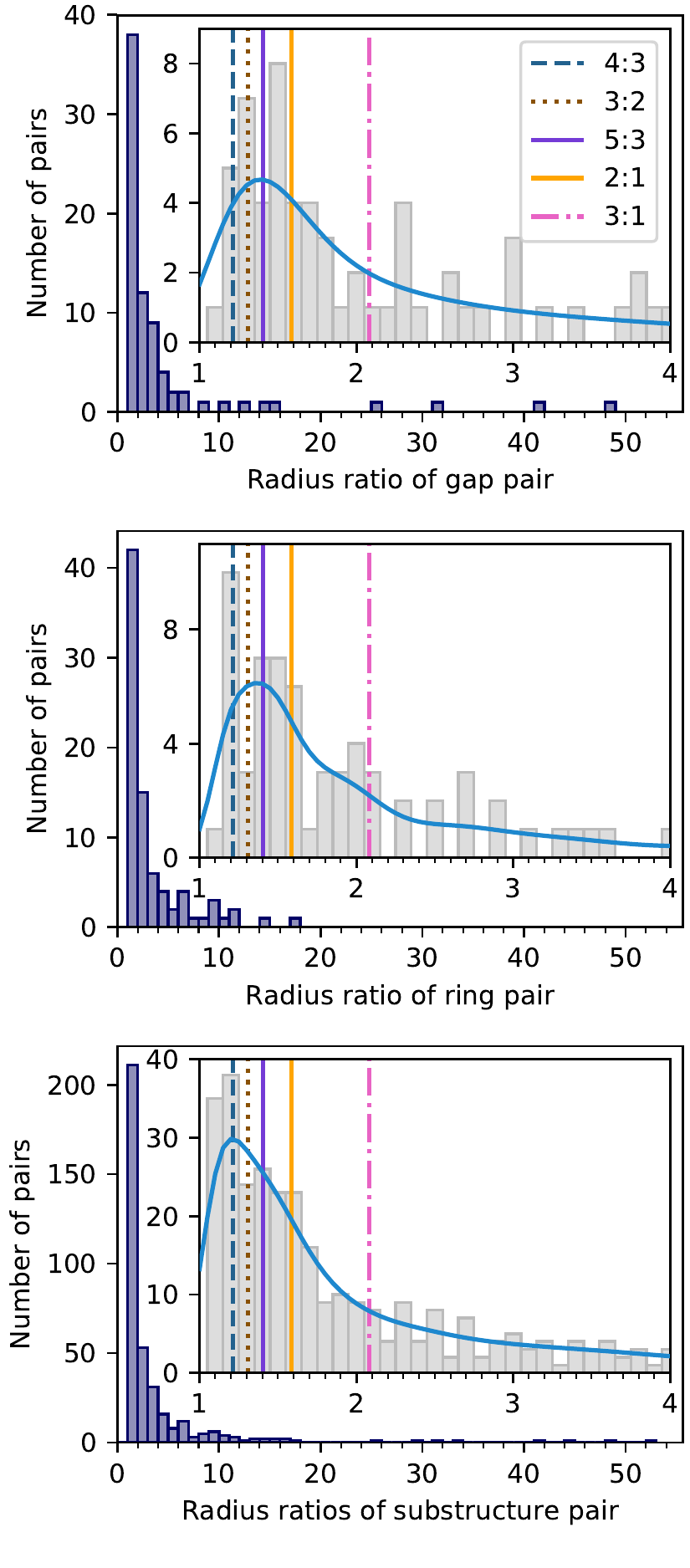}
\caption{\textit{Top:} Histogram of the distribution of the position ratios of all pairs of gaps within the same disk. Inset shows the distribution of position ratios between 1 and 4. The Gaussian kernel density estimate (scaled to match the histogram area) is shown in blue. Position ratios corresponding to low-order mean-motion resonances are marked with vertical lines. \textit{Center}: Same as above, but for pairs of rings. \textit{Bottom}: Same as above, but for all pairs of annular substructures (i.e., pairs of gaps, pairs of rings, and pairs of gaps and rings). \label{fig:positionratios}}
\end{figure}

Figure \ref{fig:positionratios} plots the distribution of the radius ratios of substructures in the DSHARP sample as well as the HL Tau and TW Hya disks. Because of the large concentration of low ratios and the sparse population of very large ratios, Gaussian kernel density estimates are derived for the distribution of log-ratios rather than for the ratios. Leave-one-out cross-validation is again used to optimize the bandwidth. The estimates are then transformed back to ratio distributions. The plots for pairs of gaps and for pairs of rings (including non-adjacent pairs) only include disks that have multiple gaps or multiple rings (i.e., excluding disks such as SR 4). The plot for pairs of all substructures includes pairs consisting of one gap and one ring, so it is not simply a combination of the two plots above. 

Within individual disks, the spread in the locations of substructures can be substantial. The innermost and outermost substructures in disks can be separated by more than 100 au, translating to position ratios exceeding 10. Overall, though, the distributions are dominated by ratios less than 2. Very low ratios ($<1.1$) are not often observed, which is probably at least partly due to angular resolution limits. 

Assuming Keplerian rotation and that the disk mass is small compared to the stellar mass, semi-major axis ratios corresponding to low-order mean-motion resonances are overplotted in Figure \ref{fig:positionratios}. There is no significant clustering around mean-motion resonances, although some pairs lie near these values. A subset of pairs that are close to resonant locations are identified in Table \ref{tab:resonances}. We list pairs with ratios within $1\%$ of the exact value, excluding substructures with positions that were estimated visually since the uncertainties are large. The B65:B41 pair from HD 143006 is the most intriguing candidate because it corresponds to two adjacent, narrow, and high-contrast rings that are very close to a 2:1 resonance. \citet{2018AA...610A..24F} previously suggested a 2:1 resonance that corresponds to B97 and D61 in the AS 209 disk, although D90-B97-D105 were thought to be a single large gap at the time due to the lower angular resolution of their observations. Multiple candidate resonances are identified for the AS 209 and HD 163296 disks, but they may arise coincidentally due to the large number of annular substructures present in these sources. On the flip side, the radial positions of shallow substructures in disks such as MY Lup and HD 142666 have large uncertainties, so it is difficult to ascertain how well their relative positions correspond to expected resonances. In addition, \citet{2015ApJ...805..100T} point out that pericenter precession can significantly shift resonant location ratios in massive disk systems. Thus, Table \ref{tab:resonances} is not exhaustive. 

One might also expect substructure pair radius ratios to cluster around characteristic values if the substructures trace molecular snowlines. Assuming a standard power law expression for the midplane temperature profile, $T_\text{mid}(r) = T_0 (r/r_0)^{-q}$ (note that Equation \ref{eq:temperature} can be re-formulated in this way), then the radius ratio corresponding to the snowlines of species $i$ and $j$ is

\begin{equation}
\frac{r_i}{r_j} = \left(\frac{T_\text{frz,j}}{T_\text{frz,i}}\right)^{1/q},
\end{equation}
where $T_\text{frz}$ denotes the freezeout temperature. While the analysis in Section \ref{sec:temperatures} is subject to large uncertainties in both the thermal profile and the gas number densities, the ratios are more robust to assumptions about these quantities. Using a representative value of $q = 0.5$ (i.e., the value corresponding to the thermal profiles used throughout this work), then $\frac{r_{\text{CO}}}{r_{\text{CO}_2}}\sim5$ if we adopt the binding energy for CO on water ice and $\sim9$ using the binding energy for CO on pure ice. As shown in Figure \ref{fig:positionratios}, few substructure pairs have these radius ratios. On the other hand, $\frac{r_{\text{N}_2}}{r_{\text{CO}}}\sim1.3$ using the binding energy for CO on pure ice and $\sim2.3$ using the binding energy for CO on water ice. A number of substructure pairs have radius ratios between these two values, but not all of them are likely to be associated with the CO and N$_2$ snowlines. First, many of the DSHARP disks have multiple pairs of close substructures. Furthermore, based on ``chemical imaging'' techniques, the CO snowline has been estimated to lie at $\sim20-30$ au \citep{2013Sci...341..630Q, 2017AA...599A.101V} in the TW Hya disk and 75 au in the HD 163296 disk \citep{2015ApJ...813..128Q}. These two sources bracket the stellar luminosities of the DSHARP sources, which suggests that substructure pairs at radii much smaller than 20 au or much larger than 75 au are also unlikely to be associated with the CO and N$_2$ snowlines.

 \begin{deluxetable}{ccccc}
\tablecaption{Candidate substructure pairs in resonance \label{tab:resonances}}
\tablehead{
\colhead{$m:n$} &\colhead{Theoretical $\frac{a_m}{a_n}$}&\colhead{Source}&\colhead{Features}&\colhead{Measured ratio}}
\startdata
2:1 &1.587&HD 143006&B65:B41&$1.591\pm0.011$\\
 & & AS 209 & B97:D61&$1.590\pm0.011$\\
 3:1 & 2.080 & HD 163296 & B100:D48 &$ 2.08\pm0.04$\\
 3:2 & 1.310& AS 209 & D137:D105 &$1.303 \pm0.011$\\
 &&Elias 20&D33:D25 & $1.31\pm0.04$\\
4:3 &1.211&AS 209 & B74:D61 & $1.220\pm0.008$ \\
&& AS 209 & D90:B74 & $1.212\pm0.006$\\
 && GW Lup &D103:B85&$1.206\pm0.012$\\
 &&RU Lup & D29:B24&$1.22\pm0.05$\\
 5:3 & 1.406& AS 209&B39:B28&$1.39\pm0.02$ \\
 && AS 209&D137:B97 & $1.417\pm0.012$\\
 &&DoAr 25 & B137:D98 & $1.40\pm0.01$\\
  &&HD 163296 &B14:D10&$1.40\pm0.17$ \\
  && HD 163296 & B67:D48 & $1.40\pm0.03$\\
  && RU Lup & B34:B24&$1.42\pm0.06$\\
 \enddata
\end{deluxetable}

\begin{figure*}[htp]
\centering
\includegraphics{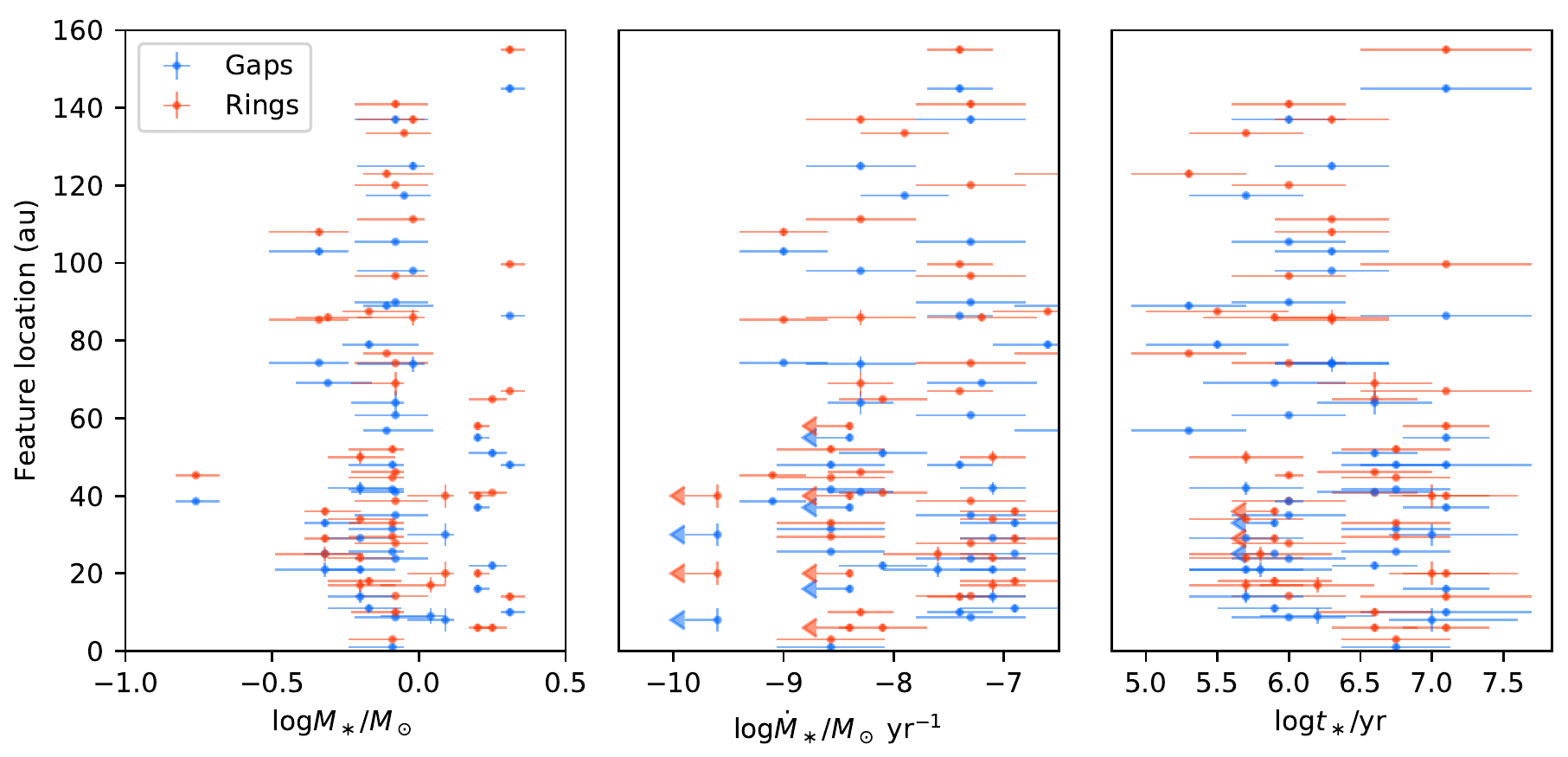}
\caption{Scatterplots of the radial positions of substructures as a function of stellar mass ($M_\ast$), stellar mass accretion rate ($\dot M_\ast$), and stellar age ($t_\ast$), respectively. $1\sigma$ error bars are shown for the stellar parameters and for the feature locations in the disk. Upper limits are shown as arrows.   \label{fig:stellarparameters}}
\end{figure*}
\subsection{Relationship with stellar parameters}

The radial positions of the substructures of the DSHARP sample and TW Hya are plotted as a function of stellar mass ($M_\ast$), stellar age ($t_\ast$), and stellar mass accretion rate ($\dot M_\ast$) in Figure \ref{fig:stellarparameters}. The stellar parameters for the DSHARP sample are taken from Table 1 of \citet{2018ApJ...869L..41A}. The adopted parameters for TW Hya are $\log M_\ast/M_\odot = -0.09\substack{+0.04\\-0.15}$, $\log t_\ast$/yr = $6.8\pm0.4$, and $\log \dot M_\ast/M_\odot$ yr$^{-1}=-8.6\pm0.5$ \citep{2018ApJ...865..157A}.  HL Tau is excluded from these plots due to the high uncertainty in its stellar properties \citep[e.g.,][]{2007ApJS..169..328R}. The uncertainties for the positions of substructures not corresponding to well-defined local maxima or minima in the radial intensity profile are estimated as the standard deviation corresponding to the major axis of the synthesized beam. No obvious trend emerges with these stellar properties, although the plots do demonstrate that annular substructures are present in disks across a large region of parameter space. The absence of an obvious trend, though, could be a consequence of the selection for bright continuum sources swamping out substructure variations with stellar properties.  

\section{Discussion \label{sec:discussion}}
\subsection{Millimeter versus scattered light substructures in the DSHARP sources}
Recent high angular resolution scattered light observations suggest that annular substructures are common not just in the disk midplane, but also the surface layers. High resolution, high sensitivity scattered light observations have been published for six of the DSHARP targets. Many come from the DARTTS-S survey by \citet{2018ApJ...863...44A}, which used SPHERE/IRDIS to observe in $J$ and $H$ band. 

In some cases, the scattered light substructures appear in close proximity to the radial locations of the millimeter continuum substructures. From SPHERE scattered light observations of HD 163296 in $J$ and $H$ band, \citet{2018AA...614A..24M} resolve a ring that matches the location of B67. SPHERE observations in $J$ band of HD 143006 reveal substructures that appear to correspond to D22, B41, D51, and B65 \citep{2018AA...619A.171B}. In SPHERE observations of AS 209, \citet{2018ApJ...863...44A} identify possible annular substructures close to B74 and B120. They also identify narrow, ringlike structures in the inner disk that are ascribed to PSF artifacts, but the presence of narrow ringlike structures in this region in the millimeter continuum raises the question of whether there are also genuine scattered light features blended in with the artifacts. A series of rings are identified in scattered light imaging of IM Lup, although \citet{2018ApJ...863...44A} comment that some of the structure could instead be tightly wrapped spiral arms. The scattered light ring located at $0\farcs58$ (92 au) is close to the apparent end of the millimeter continuum spiral arms. The second scattered light ring occurs at $0\farcs96$ (152 au), which lies just outside B134.

In other cases, scattered light substructures have no obvious correspondence to millimeter continuum substructures. \citet{2018ApJ...863...44A} detect a scattered light ring structure in the IM Lup disk at a radius of $1\farcs52$ (240 au), but there is no nearby substructure seen in the DSHARP observations. However, the SNR is low in the outer disk and the $uv$ coverage is not optimal for recovering faint, diffuse emission. The scattered light ring at $r=2\farcs1$ (332 au) lies outside the detected extent of the millimeter continuum emission. Likewise, \citet{2018ApJ...863...44A} identify a ring in scattered light at a radius of $0\farcs77$ (120 au) in the MY Lup disk, which lies outside the detected extent of millimeter continuum emission. Interestingly, although the RU Lup millimeter continuum is highly structured, the disk appears featureless in scattered light \citep{2018ApJ...863...44A}. 

\subsection{Comparison of the DSHARP sources to millimeter continuum substructures in other disks\label{sec:otherdisks}}
Besides HL Tau, TW Hya, and the DSHARP sources,  annular substructures have been identified in the millimeter continuum of at least 23 other single-disk systems hosted by Class II stars (see Table \ref{tab:otherdisks}). The list above only includes transition disks if substructures besides the large central cavity have been identified in the millimeter continuum. Otherwise, we refer the reader to \citet{2018ApJ...854..177V} for a compilation of all transition disks imaged at millimeter/sub-millimeter wavelengths up to the end of 2017 and to \citet{2018ApJ...869...17L} and \citet{2019MNRAS.482..698C} for more recent transition disk discoveries.

All spectral types from M through A are represented among disks with substructures. Among the disks observed at millimeter wavelengths prior to DSHARP, annular substructures occur far more frequently than spiral arms or azimuthal asymmetries. The DSHARP sample conforms to this trend.  Three sources (WaOph 6, IM Lup, and Elias 27) have spiral arms (see \citet{2018ApJ...869L..43H} for specific discussion of the spirals). Two sources, HD 143006 and HD 163296, have crescent-shaped azimuthal asymmetries near well-defined ring substructures. Outside the DSHARP sample, several other disks have also been shown to exhibit annular substructure in conjunction with spiral arms and azimuthal asymmetries \citep[e.g.,][]{2018ApJ...860..124D,2018AA...619A.161C}. Previous observations and DSHARP together suggest a trend of high-contrast azimuthal asymmetries (i.e., azimuthal contrasts larger than a factor of a few) accompanying ring structures preferentially in disks hosted by stars with spectral types of G and earlier.

Because most disks outside DSHARP have been observed at lower angular resolution, detections of disk substructures have been biased toward high-contrast gaps and rings with widths of tens of au. The DSHARP observations, as well as a recent Taurus survey at a spatial resolution of $\sim15$ au by \citet{2018ApJ...869...17L}, suggest that narrower substructures are more typical. Higher-resolution observations will likely reveal additional annular substructures in the other disks listed in Table \ref{tab:otherdisks}. 

While DSHARP excluded large-cavity transition disks, many have now been observed at moderate-to-high angular resolution (see Table \ref{tab:otherdisks}). Large-cavity transition disks have often been analyzed as a group distinct from ``full'' disks \citep[e.g.,][]{2018ApJ...859...32P, 2018ApJ...854..177V}, but the distinction between the two categories ironically becomes blurrier at higher angular resolutions.  Substructures detected in ``transition'' disks exhibit a level of diversity similar to the DSHARP sample. Observed features also include annular gaps and rings, spiral arms, and azimuthal asymmetries \citep[e.g.,][]{2017AA...600A..72F, 2018ApJ...860..124D}. Millimeter continuum observations have also begun to detect small inner disks within transition disk cavities \citep[e.g.,][]{2017ApJ...840...60B, 2018ApJ...860..124D, 2018ApJ...868L...5K}, suggesting that at least some transition disk cavities could be considered (particularly wide) annular gaps. Meanwhile, several of the DSHARP disks  turned out to have small central emission deficits. In addition, while the millimeter continuum emission of the SR 4 disk is centrally peaked, it features a deep annular gap in the inner disk and the SED shows a near-infrared deficit that is reminiscent of transition disk SEDs \citep{2018ApJ...869L..41A}. It is not clear whether these various types of sources represent different stages along the same evolutionary path, but they at least suggest that the ``transition''  vs. ``full'' disk dichotomy is too simplistic. 

\begin{deluxetable*}{ccccc}
\tablecaption{Other disks with millimeter continuum annular substructures \label{tab:otherdisks}}
\tablehead{\colhead{Source} &\colhead{SpT} &\colhead{ Ref.\tablenotemark{a}} &$\theta_\text{b,min }\times d$ (au)\tablenotemark{b}& \colhead{Notes on other millimeter features}}
\startdata
HD 97048 &A0-B9 &1,2 &48&Classified as transition disk\\
MWC 480 & A5 &3,4 &18&\\
HD 169142 & A5 &5,6&16&Classified as transition disk\\
MWC 758 &A8 &7,8&6.2&Classified as transition disk; high-contrast asymmetry; spiral arm(s)\\
V1247 Ori & F0 & 9,10&13&Classified as transition disk; high-contrast asymmetry\\
HD 142527 &F6 & 11,12&20&Classified as transition disk; high-contrast asymmetry; circumbinary disk\\
SAO 206462 & F8 &13,14&8.5&Classified as transition disk; high-contrast asymmetry \\
RY Tau & G0 &15,4&14& Classified as transition disk; high-contrast asymmetry\\
CI Tau &K5.5 &15,16&4.7&\\
DL Tau & K5.5 & 15,4&17&\\
GM Aur & K6 &15,17&32&Classified as transition disk \\
V1094 Sco & K6 &18,19&28& \\
PDS 70 & K7 &20,21&19&Classified as transition disk \\
Sz 98 & K7 & 18, 22&39&\\
DN Tau & M0.3 & 15,4&14\\
DS Tau & M0.4&15,4&16&\\
AA Tau &M0.6&15,23&15& \\
IQ Tau & M1.1&15,4&14&\\
UZ Tau E & M1.9 &15,4&14& \\
GO Tau & M2.3&15,4&16&\\
FT Tau & M2.8&15,4&14&\\
DM Tau & M3.0 & 15,24&4.5&Classified as transition disk \\
WSB 82 & $-$ & $-$,25 & 28&Classified as transition disk \\
\enddata
\tablerefs{(1) \citealt{1987MNRAS.224..497W} (2) \citealt{2017AA...597A..32V} (3) \citealt{2001AA...378..116M}(4) \citealt{2018ApJ...869...17L} (5) \citealt{1997MNRAS.286..604D} (6) \citealt{2017AA...600A..72F} (7) \citealt{1999AA...343..163B} (8) \citealt{2018ApJ...860..124D} (9) \citealt{2003AJ....126.2971V} (10) \citealt{2017ApJ...848L..11K} (11) \citealt{2018AA...617A..37C} (12) \citealt{2018ApJ...864...81O} (13) \citealt{1995MNRAS.274..977C} (14) \citealt{2018AA...619A.161C} (15) \citealt{2014ApJ...786...97H} (16) \citealt{2018ApJ...866L...6C} (17) \citealt{2018ApJ...865...37M} (18) \citealt{alcala17} (19) \citealt{2018AA...616A..88V} (20) \citealt{2016MNRAS.461..794P} (21) \citealt{2018ApJ...858..112L} (22) \citealt{2017AA...606A..88T} (23) \citealt{2017ApJ...840...23L} (24) \citealt{2018ApJ...868L...5K} (24) \citealt{2017ApJ...851...83C}}
\tablenotetext{a}{First reference is for spectral type, second reference is for millimeter continuum observations of annular substructures}
\tablenotetext{b}{Distances computed using \textit{Gaia} parallaxes \citep{2018AA...616A...1G}}
\end{deluxetable*}

\subsection{The prevalence of millimeter continuum substructures}
\subsubsection{Prevalence in Ophiuchus and Lupus}
The widespread presence of annular substructures in the DSHARP sample naturally raises the question of whether such structures are ubiquitous. Since DSHARP selected for bright disks, it is not straightforward to derive an occurrence rate for millimeter continuum substructures. Nevertheless, interesting lower bounds can be set for the Lupus and Ophiuchus star-forming regions, from which the bulk of the DSHARP sample originates. 

The Lupus star-forming region is $\sim1-3$ Myr old and lies $\sim150-170$ pc away \citep[e.g.,][]{2008ApJS..177..551M, 2014AA...561A...2A,2016AA...595A...2G,2018AA...616A...1G}. Based on optical and IR spectra, 96 Class II or flat spectrum protoplanetary disks hosted by stars with $M_\ast>0.1$ $M_\odot$ were identified in the Lupus I-IV clouds and subsequently targeted at millimeter/sub-millimeter wavelengths at an angular resolution of $\sim0\farcs3$ \citep{2016ApJ...828...46A, 2018ApJ...859...21A}. Of the 71 disks with millimeter detections, at least 8 exhibit annular substructure: GW Lup, IM Lup, RU Lup, Sz 114, Sz 129, MY Lup, HK Lup, and V1094 Sco \citep[this work,][]{2017AA...606A..88T, 2018AA...616A..88V}. Thus the fraction of Class II Lupus disks with annular substructure is $\gtrapprox 10\%$. This lower bound is extremely conservative, considering that the majority of the substructures observed in the Lupus disks are not evident at the angular resolution of the Lupus survey. For perspective, though, 10 large-cavity transition disks have been identified in Lupus \citep{2018ApJ...854..177V}. If considered as distinct categories, disks with annular substructure cannot be much rarer than large-cavity transition disks. Taken together, disks with some kind of substructure comprise at least a quarter of all Class II Lupus disks detected at millimeter wavelengths. Moreover, at least among the bright disks, substructures are almost always present. Of the 15 Lupus disks with 1.3 mm continuum fluxes equal to or brighter than the faintest DSHARP disk (DoAr 33), 12 are known so far to have either cavities or annular substructures. Another one of the 15, HT Lup, has spiral arms in its primary disk but does not exhibit annular substructure at the angular resolution of DSHARP \citep{2018ApJ...869L..44K}. 

The c2d Spitzer Legacy project identified 223 flat spectrum and Class II disks in the Ophiuchus star-forming region, which is similar in age to Lupus but features a higher star formation rate and stellar number density \citep[e.g.,][]{2009ApJS..181..321E}. The largest millimeter continuum survey so far of Ophiuchus disks, ODISEA, targeted $\sim$130 Class II and flat spectrum disks (and a small number of Class I disks) with ALMA at an angular resolution of $\sim0\farcs2$ \citep{2019MNRAS.482..698C}.  At least 7 Class II Oph disks are known to have annular substructures: SR 4, Elias 20, DoAr 25, Elias 24, Elias 27, DoAr 33, and WSB 82 \citep[this work,][]{2017ApJ...851...83C}, along with an 8th tentative detection in WSB 52. WaOph 6 and AS 209 also have substructures but are only sometimes classified as Ophiuchus sources. Twenty-seven of the flat spectrum/Class II disks in ODISEA have millimeter continuum fluxes equal to or brighter than the faintest DSHARP disk (including DoAr 33 itself). Thirteen of these are known so far to have some kind of millimeter continuum substructure, including cavities and annular substructures. Again, it should be emphasized that the other sources were observed at much coarser angular resolution than the DSHARP sample, but even the existing observations suggest that the prevalence of millimeter continuum substructures is high among bright disks at least. The  angular resolution of ODISEA is sufficient to show, though, that it is not common for the Ophiuchus disks to have the kind of deep, wide gaps observed in disks around sources such as AS 209, Elias 24, HD 163296, and HD 143006. 
 
\subsubsection{Small versus large disks}
Although the radial extents of the DSHARP disks span an order of magnitude, the selection criteria created a sample in which large disks are overrepresented \citep{2018ApJ...869L..41A}. The DSHARP sources are larger than about 40$\%$ of the 105 disk targets drawn from various star-forming regions \citep{2017ApJ...845...44T, 2018ApJ...865..157A} and about three-quarters of the flat spectrum/Class II disks in Ophiuchus \citep{2019MNRAS.482..698C}. Likewise, other disks with annular substructures reported in millimeter continuum emission (see Section \ref{sec:otherdisks}) have sizes similar to the DSHARP sources.

SR 4, WSB 52, and DoAr 33 are the smallest of the DSHARP sources as traced by millimeter continuum emission. While the gap in SR 4 has one of the highest contrasts measured for the DSHARP sources, DoAr 33 and WSB 52 have very low-contrast substructures compared to the larger DSHARP disks. Furthermore, as shown in Figure \ref{fig:positionhistogram}, a comparatively small number of annular substructures are detected in the inner 40 au of the DSHARP disks.  In a separate survey, \citet{2018ApJ...869...17L} find indications that larger disks tend to have a larger number of rings. This raises the question of whether the high prevalence of annular substructures in the DSHARP sample signals ubiquity in the overall disk population, or just among larger disks. It has long been a puzzle how large protoplanetary disks persist for Myr timescales when radial drift of millimeter-size grains is expected to act quickly \citep[e.g.,][]{2002ApJ...581.1344T,2007AA...469.1169B}. A possible solution is that disk substructures impede radial drift, allowing large disk sizes to be maintained \citep[e.g.,][]{1972fpp..conf..211W, 2012AA...538A.114P}.  

Surveys imaging smaller disks will be essential to determine the extent to which disks of more typical sizes resemble the DSHARP sources. However, if the small disks are also optically thick out to radii of $r\sim10$ au, then emission substructures may not be observable even if surface density variations are present. Thus, non-detections of substructures in small disks will need to be interpreted carefully.

\subsubsection{Relationship with age}
The median age of the DSHARP sources is $\sim1$ Myr, and the youngest sources have estimated ages of only a few hundred thousand years (with the caveat that age estimates have large uncertainties). The high prevalence of annular substructures suggests that they form relatively quickly in disks. Given how readily substructures have been detected in Class II disks, a natural question is whether some or all of these structures actually formed while the sources were still embedded. Substructures have been reported in a few Class I (or borderline Class I/Class II) disks \citep{2015ApJ...808L...3A, 2016Natur.535..258C,2018ApJ...857...18S}, but a systematic high angular resolution study of Class I disks will be necessary to establish whether these substructures are common at this earlier stage. 

Not many disks older than a few Myr have been observed at high angular resolution, but several older disks (e.g., TW Hya, HD 143006, HD 142666, and HD 163296) are known to have annular substructure. This suggests either that these structures are long-lived in disks or that their formation can take place at very different points in the lifetime of a disk. Other than TW Hya, though, the older sources have spectral types of G or earlier and the younger sources are K or M stars, which makes it difficult to isolate trends with disk age.

\subsection{Possible origins of annular substructures}
\subsubsection{Planets}
The most popular hypothesis for the origin of annular substructures in disks is that they result from gravitational interactions between the disk and one or more planets \citep[e.g.,][]{1984ApJ...285..818P, 2004AA...425L...9P,2010AA...518A..16F}. Of the DSHARP sources, the companion hypothesis has specifically been explored for the AS 209, Elias 24, Elias 27, and HD 163296 disk via hydrodynamical simulations attempting to reproduce previous millimeter continuum observations \citep[e.g][]{2016PhRvL.117y1101I, 2018AA...610A..24F, 2018ApJ...860L..12T, 2018MNRAS.475.5296D, 2018ApJ...860L..13P, 2017ApJ...839L..24M,2018ApJ...860L...5F,2018ApJ...866..110D}, although models of the Elias 27 disk have focused on the spiral arms rather than the annular substructures. Planet-disk interactions involving protoplanets ranging from Saturn to Jupiter mass can reproduce the lower-resolution observations, but the detection of additional substructures in the DSHARP observations makes these inferences worth revisiting. \citet{2018ApJ...869L..47Z} perform a parameter space study to estimate the planet masses compatible with the gaps observed in the DSHARP sample. 

Although large masses have been inferred for many candidate protoplanets in disks with substructure, recent works have explored the possibility that these substructures are created instead by lower-mass planets (i.e., super-Earths) in low-viscosity disks \citep[e.g.,][]{2013ApJ...769...41D, 2014ApJ...785..122Z, 2017ApJ...843..127D, 2017ApJ...850..201B}. Whereas each gap has to be opened by a massive planet in higher-viscosity disks, a single planet can open multiple gaps in low-viscosity disks. In particular, \citet{2014ApJ...785..122Z} and \citet{2017ApJ...843..127D} demonstrate that a super-Earth can create a ``double-gap'' feature wherein annular gaps open interior and exterior to its orbit at a distance of about one pressure scale height.  If $h/r \sim 0.1$, then potential ``double gap'' features among DSHARP sources include D90:D105 in AS 209, D74:D98 in DoAr 25, D25:D33 in Elias 20, and D21:D29 in RU Lup. What is particularly interesting about these pairs is that they create ``w-features'' in the radial intensity profile\textemdash that is, the intensity just interior and exterior to the double gap region is comparable to or brighter than the ring separating the two gaps, forming a section in the radial profile that looks like the letter ``w.'' This is not in general a trait of an arbitrary pair of consecutive gaps in the DSHARP sample, but it \textit{is} a predicted characteristic of double gaps opened by a planet in a low-viscosity disk. 

Hydrodynamical simulations of planet-disk interactions indicate that the eccentricity of annular substructures increases for more massive planets and for lower disk viscosities \citep[e.g.][]{1991ApJ...381..259L,2006AA...447..369K,2018ApJ...869L..47Z}. Several disks have been proposed to be eccentric based either on offsets between the measured centers of various features \citep{2015ApJ...808L...3A, 2018ApJ...860..124D} or residuals from subtracting an axisymmetric model \citep{2016PhRvL.117y1101I, 2018ApJ...869L..49I}. For the most part, the DSHARP substructures appear to be concentric within the stated uncertainties of the fitted ellipse center positions. The $1\sigma$ uncertainties in $\Delta x$ and $\Delta y$ are typically $1-3$ mas each, translating to a few tenths of an au at a typical distance of 140 au. For a face-on disk with well-resolved substructure, offsets of $\sim1$ au between the centers of substructures should be detectable. If taken to represent the distance between the center of an eccentric substructure and the star at one focus, this offset would correspond to an eccentricity of $\sim0.05$ at $r=20$ au and $\sim0.01$ at $r=100$ au. In the AS 209 disk, the measured center of B120 appears offset from B74 by $6.5\pm1.4$ mas ($0.79\pm0.17$ au), which might suggest a very small eccentricity of $\sim0.01$ at $r=120$ au. In practice, it is not straightforward to translate measured offsets to eccentricities because vertical structure and optical depth complicate the apparent offsets of substructures in inclined disks \citep[e.g.,][]{2016ApJ...816...25P}, and substructures with the same eccentricity will not be offset from one another. Forward radiative transfer modeling will be useful to search for more subtle signatures of eccentricity. The inner rings of Sz 129 and HD 142666 may be worthy of further investigation due to the asymmetries noted in Section \ref{sec:additionalstructures}.  \citet{2018ApJ...869L..49I} also discuss possible eccentricity in the HD 163296 disk.

Of the substructures in the DSHARP targets, the ones in the AS 209 and HD 163296 disk may present the most intriguing case so far for originating from planet-disk interactions. Substructures that have been observed in CO isotopologue emission \citep{2016ApJ...823L..18H, 2016PhRvL.117y1101I} may be indicative of gas surface density substructures created by a perturber. \citet{2018ApJ...860L..12T} also find evidence of non-Keplerian gas motion coinciding with the millimeter continuum substructures in the HD 163296 disk. In addition, \citet{2018MNRAS.479.1505G} reported the detection of a point source in $L'$ band within the D48 feature of the HD 163296 disk, although further observations will be necessary to establish whether the source is a protoplanet. Further arguments for the planet-disk interactions scenario for these two disks are discussed in \citet{2018ApJ...869L..48G}, \citet{2018ApJ...869L..49I}, and \citet{2018ApJ...869L..47Z}.

\subsubsection{Snowlines}

Previously detected substructures in the HL Tau, TW Hya, and HD 163296 disks have been hypothesized to be due to volatile freezeout altering the coagulation and fragmentation properties of dust grains \citep[e.g.,][]{2015ApJ...806L...7Z, 2016ApJ...821...82O,2017ApJ...845...68P}. Within the DSHARP sample, the evidence does not seem to point to a snowline origin for most of the substructures. First, since the stellar luminosity plays a central role in setting the thermal profile of the disk, the radii of snowline-related substructures would be expected to scale with luminosity. However, as shown in Figure \ref{fig:linearprofiles} and Figure \ref{fig:snowlines}, there is not an obvious correspondence between the stellar luminosities/thermal profiles and the location of substructures. In addition, the wide variations in the number of substructures identified in each disk would seemingly require that the major volatile species differ from disk to disk. Furthermore, as discussed in Section \ref{sec:spacings}, the relative spacings of the substructures do not suggest that they are primarily associated with snowlines. Of course, if only a single substructure in each disk is related to a snowline, the relative spacings would not follow a pattern. However, since most disks exhibit multiple annular substructures, this would still mean that the majority of the substructures are unrelated to snowlines. 

On the other hand, ascertaining whether an individual substructure is related to a snowline is challenging. To help distinguish between snowline and planetary origins, \citet{2017ApJ...845...68P} suggest comparing the millimeter emission profiles to scattered light emission profiles\textemdash snowlines should not alter the gas surface density and would therefore lead to scattered light substructures that are wider and higher contrast than the millimeter substructures. The emission gaps appear to be deeper and wider at millimeter wavelengths for the AS 209, HD 143006, MY Lup, and RU Lup disks compared to scattered light observations, contrary to expectations for snowline-created structure. In a complementary analysis of several DSHARP sources, \citet{2018ApJ...869L..46D} find evidence for dust trapping in the high-contrast outer rings in the AS 209, HD 163296, GW Lup, and Elias 24 disks based on comparisons between the ring widths and pressure scale heights. This provides a further argument against the snowline origin, which is not expected to create the requisite gas pressure bumps to trap dust. Finally, while the annular substructures appear relatively shallow and narrow in both scattered light and millimeter continuum emission for the IM Lup disk, \citet{2018ApJ...865..155C} predict that the CO snowline is located at $r\sim40$ au, whereas the annular substructures in the IM Lup disk are detected well beyond $r=100$ au. Even given the large uncertainties associated with the thermal profile, the IM Lup substructures appear to be much too distant from the central star to be associated with a major volatile snowline. 

The lack of an obvious association between substructure and expected snowline locations is in line with several other studies of disks. In a re-analysis of the thermal structure of the HL Tau disk based on modeling multi-band high resolution ALMA images, \citet{2016ApJ...816...25P} find that the disk substructures do not match with predicted snowline locations. \citet{2018AA...616A..88V} show that the annular substructures in the V1094 Sco disk occur at temperatures much lower than the freezeout temperatures of major disk volatiles. Most recently, in an analysis of 12 disks in Taurus, \citet{2018ApJ...869...17L} find that annular substructures have a large spread in radial locations and do not occur preferentially near snowlines.

\subsubsection{The role of internal gas dynamics}
Internal gas dynamics arising from the coupling between the magnetic field and the disk dust and gas may also give rise to annular substructure formation. Zonal flows due to MRI turbulence or spontaneous concentrations of magnetic flux can lead to gas density variations up to a few tens of percent across radial extents of several pressure scale heights \citep[e.g.,][]{2009ApJ...697.1269J,2014ApJ...796...31B,2014ApJ...784...15S}, but there remain large uncertainties associated with the global magnetic flux evolution \citep[e.g.,][]{1994MNRAS.267..235L,2014ApJ...785..127O,2014MNRAS.441..852G,2017ApJ...836...46B}. An additional source of uncertainty is the extent of particle trapping within such gas substructures, so it is not yet clear whether the extremely high contrast continuum substructures observed in a few of the DSHARP sources (AS 209, HD 143006, HD 163296, Elias 24) can be created through these ``magnetic'' mechanisms. In the dead zone scenario, the locations of annular substructures in the dust and gas are associated with a transition in the disk's radial resistivity profile that should occur well under a radius of 100 au \citep[e.g.,][]{2010AA...515A..70D, 2015AA...574A..68F, 2015AA...574A..10L, 2016AA...590A..17R}. The majority of the DSHARP substructures occur within $r=100$ au, but a third of these disks have substructures that lie outside the largest expected radial extent of a dead zone. Moreover, \citet{2016AA...590A..17R} comment that vortices appear along with the ring structures in the majority of their simulations, whereas most of the DSHARP disks do not have obvious asymmetric features indicative of vortices. 

Absent a direct detection of a planet, it is not observationally straightforward to distinguish between planet-disk interactions and magnetohydrodynamic effects \citep[e.g.,][]{2015AA...574A..68F, 2016AA...590A..17R}. Studies of molecular lines at high angular and/or spectral resolution could help to address some key uncertainties, with some caveats. First, the strength and morphology of magnetic fields in disks remain unknown, although future measurements of Zeeman splitting could shed some light \citep[e.g.,][]{2017AA...607A.104B}.  Efforts have also been made to constrain the disk ionization fraction via observations of molecular ions \citep[e.g.,][]{2011ApJ...743..152O,2015AA...574A.137T,2015ApJ...799..204C}, but identifying the boundary of the dead zone is not straightforward due to the relatively low angular resolution data available thus far and uncertainties about the abundances of charge carriers that have not been detected. Studies of turbulence via molecular line observations may also help to characterize the overall disk gas dynamics, including the extent of the dead zone, but degeneracies with the thermal profile still pose a major challenge \citep[e.g.,][]{2015ApJ...813...99F,2018ApJ...864..133T}. 

\subsubsection{Other hypotheses}
The list of hypotheses for the origins of substructures in disks is quite extensive. We briefly address a couple additional hypotheses that are also of interest given the characteristics of some of the DSHARP sources. 

X-ray photoevaporation has been proposed as a mechanism for dispersing disks from the inside-out \citep[e.g.,][]{2008ApJ...688..398E, 2010MNRAS.401.1415O, 2017MNRAS.464L..95E}. While this mechanism is not expected to generate the annular substructures observed in the DSHARP sample, it could be responsible for the small inner gaps/cavities observed in the HD 143006, HD 142666, and Sz 129 disks.

Secular gravitational instability may create annular substructures in disks with low turbulence, low gas-to-dust ratios, and Toomre $Q\lessapprox 3$  \citep[e.g.,][]{2014ApJ...794...55T}.  Few estimates exist for disk turbulence, but the turbulent line broadening upper limits derived for HD 163296 and TW Hya are less than a tenth of the sound speed, which is considered weak \citep[e.g.,][]{2015ApJ...813...99F,2018ApJ...856..117F}. \citet{2017AA...599A.113M} estimated that the gas-to-dust ratio is less than 10 for RU Lup, GW Lup, Sz 114, MY Lup, and Sz 129, which may make them interesting candidates for secular GI. However, \citet{2017AA...599A.113M} also present the alternative scenario that the disks have normal gas-to-dust ratios but are depleted in CO, yielding artificially low gas mass estimates. 

Overall, the planet-disk interaction scenario so far seems best able to accommodate the diversity of substructures in disks. Planets can form at any radius where substructures are observed. The diverse exoplanet masses measured offer a straightforward way to explain the large variations observed in gap widths and contrasts. Exoplanetary systems also contain differing numbers of planets, which can help to explain the different number of substructures observed from disk to disk. Even a given planet mass at a given radius can create vastly differing dust substructures depending on the disk viscosity and dust size distribution \citep[e.g.,][]{2018ApJ...864L..26B}. While certain individual substructures may be associated with snowlines, the snowline scenario does not seem flexible enough to explain the widely varying number and locations of substructures from disk to disk. More constraints on turbulence and magnetic field properties in disks will be needed to determine whether internal gas dynamics are sufficient to explain the presence of annular substructures without invoking planets. Direct imaging searches for planets and high angular and spectral resolution molecular line observations will provide additional key pieces of information for determining the origins of disk substructures.  

\section{Summary\label{sec:summary}}

We present a systematic analysis of annular substructures detected in high angular resolution millimeter continuum observations of 18 single-disk systems from DSHARP. The DSHARP observations suggest that protoplanetary disk substructures are common and thus must trace fundamental evolutionary processes in disks. Our findings are as follows: 
\begin{itemize}
\item Substructures are detected in the millimeter continuum of all 18 disks. Annular substructures are identified in at least 17 disks and tentatively in an eighteenth, making them far more common than spiral arms and crescent-like azimuthal asymmetries. A minority of disks also feature other kinds of substructures\textemdash spiral arms are observed in three of these sources and crescent-like azimuthal asymmetries in another two. 

\item The dust disk morphologies are diverse, with no two disks having similar emission profiles. Substructures can occur at essentially any disk radius probed by ALMA, with features detected in the inner 5 au of a disk and past radii of 150 au. DSHARP detected both shallow, tightly packed gaps and deep, well-separated gaps, sometimes within the same disk. 

\item Annular substructures are observed in disks hosted by stars across a range of luminosities,  masses, and accretion rates. No immediately obvious relationships emerge yet between the substructure locations and stellar properties, although underlying relationships may be obscured by the selection bias toward bright disks. 

\item The relative spacings of the substructures and the lack of an apparent correspondence to the thermal profiles suggest that most annular substructures do not trace the locations of snowlines, although it is possible that some individual features are associated with snowlines. 

\item The observed annular substructures are reminiscent of features observed in simulations of planet-disk interactions. In at least a few disks, some morphological characteristics are particularly suggestive, including substructures in possible mean-motion resonances and ``double gap'' features resembling hydrodynamical predictions for perturbations by a planet in a low-viscosity disk. 
\end{itemize}

\acknowledgments
This paper makes use of ALMA data\\
\dataset[ADS/JAO.ALMA\#2016.1.00484.L]{https://almascience.nrao.edu/aq/?project\_code=2016.1.00484.L},\\
\dataset[ADS/JAO.ALMA\#2013.1.00694.S]{https://almascience.nrao.edu/aq/?project\_code=2013.1.00694.S},\\
 \dataset[ADS/JAO.ALMA\#2013.1.00226.S]{https://almascience.nrao.edu/aq/?project\_code=2013.1.00226.S}, \\
\dataset[ADS/JAO.ALMA\#2013.1.00366.S]{https://almascience.nrao.edu/aq/?project\_code=2013.1.00366.S},\\
 \dataset[ADS/JAO.ALMA\#2013.1.00498.S]{https://almascience.nrao.edu/aq/?project\_code=2013.1.00498.S},\\
  \dataset[ADS/JAO.ALMA\#2013.1.00601.S]{https://almascience.nrao.edu/aq/?project\_code=2013.1.00601.S}, \\
  \dataset[ADS/JAO.ALMA\#2013.1.00798.S]{https://almascience.nrao.edu/aq/?project\_code=2013.1.00798.S},\\
   \dataset[ADS/JAO.ALMA\#2015.1.00486.S]{https://almascience.nrao.edu/aq/?project\_code=2015.1.00486.S},\\
   \dataset[ADS/JAO.ALMA\#2015.1.00964.S]{https://almascience.nrao.edu/aq/?project\_code=2015.1.00964.S}. We thank the referee for comments improving this paper and NAASC and JAO staff for their advice on data calibration and reduction. ALMA is a partnership of ESO (representing its member states), NSF (USA) and NINS (Japan), together with NRC (Canada) and NSC and ASIAA (Taiwan), in cooperation with the Republic of Chile. The Joint ALMA Observatory is operated by ESO, AUI/NRAO and NAOJ. The National Radio Astronomy Observatory is a facility of the National Science Foundation operated under cooperative agreement by Associated Universities, Inc. J.H. acknowledges support from the National Science Foundation Graduate Research Fellowship under Grant No. DGE-1144152. S.A. and J.H. acknowledge support from the National Aeronautics and Space Administration under grant No.~17-XRP17$\_$2-0012 issued through the Exoplanets Research Program.  C.P.D. acknowledges support by the German Science Foundation (DFG) Research Unit FOR 2634, grants DU 414/22-1 and DU 414/23-1. A.I. acknowledges support from the National Aeronautics and Space Administration under grant No.~NNX15AB06G issued through the Origins of Solar Systems program, and from the National Science Foundation under grant No.~AST-1715719.  L.P. acknowledges support from CONICYT project Basal AFB-170002 and from FCFM/U.~de Chile Fondo de Instalaci\'on Acad\'emica. V.V.G. and J.C acknowledge support from the National Aeronautics and Space Administration under grant No.~15XRP15$\_$20140 issued through the Exoplanets Research Program. Z.Z. and S.Z. acknowledge support from the National Aeronautics and Space Administration through the Astrophysics Theory Program with Grant No.~NNX17AK40G and the Sloan Research Fellowship. M.B. acknowledges funding from ANR of France under contract number ANR-16-CE31-0013 (Planet Forming disks). T.B. acknowledges funding from the European Research Council (ERC) under the European Union's Horizon 2020 research and innovation programme under grant agreement No.~714769.  L.R. acknowledges support from the ngVLA Community Studies program, coordinated by the National Radio Astronomy Observatory, which is a facility of the National Science Foundation operated under cooperative agreement by Associated Universities, Inc. 

\facilities{ALMA}

\software{ \texttt{analysisUtils} (\url{https://casaguides.nrao.edu/index.php/Analysis_Utilities}), \texttt{AstroPy} \citep{2013AA...558A..33A}, \texttt{CASA} \citep{2007ASPC..376..127M}, \texttt{emcee} \citep{2013PASP..125..306F}, \texttt{matplotlib} \citep{Hunter:2007}, \texttt{scikit-image} \citep{scikit-image}, \texttt{scikit-learn} \citep{scikit-learn}, \texttt{SciPy} \citep{scipy}}

\appendix
\section{Azimuthal angle convention\label{sec:angles}}
\begin{figure}[htp]
\centering
\includegraphics{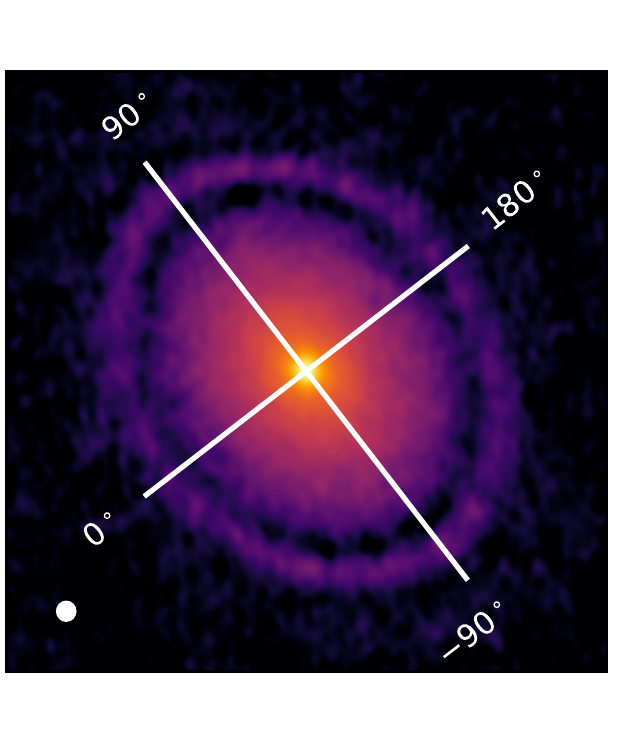}
\caption{Diagram of the azimuthal angle convention used in the DSHARP paper, demonstrated for the GW Lup disk. North is up and east is left.  \label{fig:angles}}
\end{figure}

This section describes the azimuthal angle convention used for the deprojected coordinate system throughout the DSHARP papers.  The positive $y$-axis of the deprojected coordinate system is rotated east of north by the position angle of the disk. The polar angle increases in the clockwise direction (i.e., if the position angle of a disk is 0$^\circ$, then $\theta=90^\circ$ is in the direction of increasing declination and $\theta=0^\circ$ is in the direction of increasing right ascension). This is diagrammed in Figure \ref{fig:angles} for the GW Lup disk.  

\section{Substructure width definitions \label{sec:schematic}}
\begin{figure}[htp]
\centering
\includegraphics{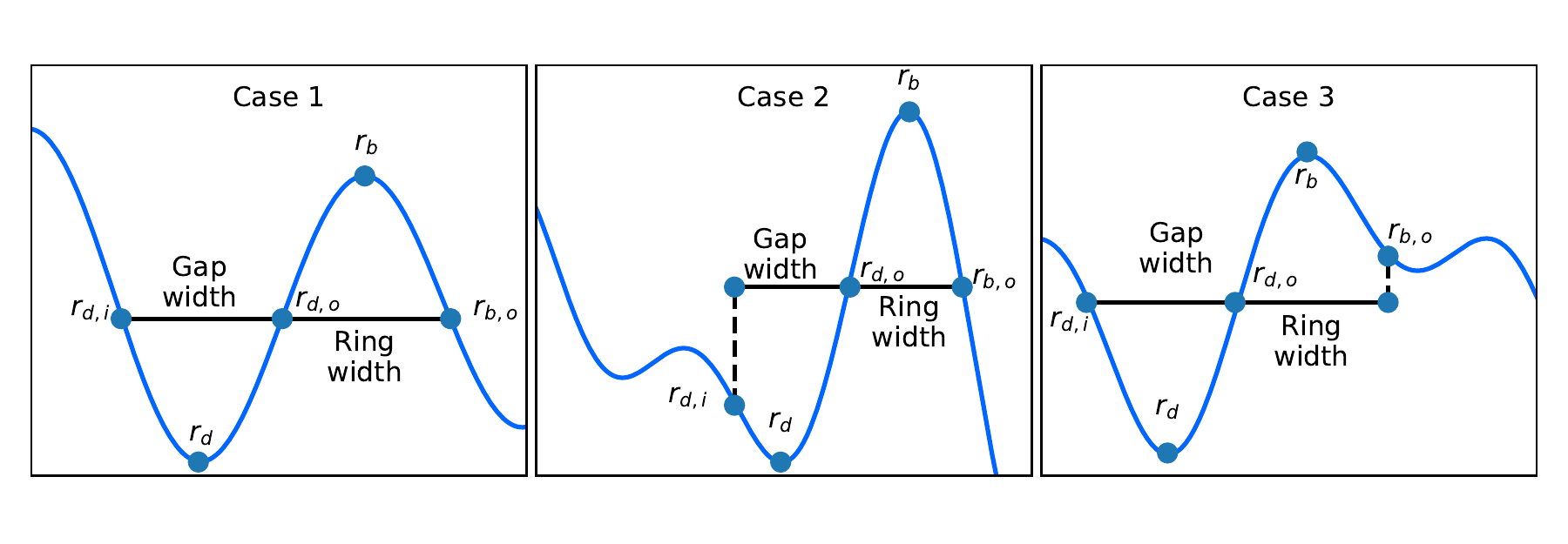}
\caption{Schematic of how quantities are defined for measurements of substructure widths in Section \ref{sec:widths}. Case 1 is the standard scenario. Case 2 represents scenarios in which a gap is exterior to a ring with a peak intensity value less than $I_\text{mean}$. Case 3 represents scenarios in which a ring is interior to a gap with a minimum intensity value greater than $I_\text{mean}$\label{fig:schematic}.}
\end{figure}

A schematic for substructure width definitions is presented in Figure \ref{fig:schematic}. Case 1 is the standard scenario described in Section \ref{sec:widths}. In Case 2, the peak of a ring immediately interior to a gap has a lower intensity value than $I_\text{mean}$, so $r_{d,i}$ is set equal to $r_{b,o}$ for the interior ring. This applies to the B97-D105-B120 sequence in the AS 209 disk and the B72-D77-B85 sequence for the HL Tau disk.  In Case 3,  the trough of a gap exterior to a ring has a higher intensity value than $I_\text{mean}$, so $r_{b,o}$ is set equal to $r_{d,i}$ for the exterior gap. This applies to the D34-B40-D44 sequence in the HL Tau disk.   

\bibliographystyle{aasjournal}

\end{document}